\documentclass[12pt,fleqn, a4paper]{article}

\newcommand{\ra}{\rightarrow}
\newcommand{\bra}{\langle} \newcommand{\ket}{\rangle}
\newcommand{\be}{\begin{equation}}
\newcommand{\ee}{\end{equation}}
\newcommand{\bea}{\begin{eqnarray}}
\newcommand{\eea}{\end{eqnarray}}
\newcommand{\eps}{\epsilon}
\newcommand{\E}{\mbox{e}}

\newcommand{\ffi}{\varphi}

\newcommand{\ep}{\qquad {\vrule height 10pt width 8pt depth 0pt}}

\newcommand{\grintl}{[\kern-.18em [}
\newcommand{\grintr}{]\kern-.18em ]}

\newtheorem{lem}{Lemma}[section]
\newtheorem{prop}{Proposition}[section]
\newtheorem{thm}{Theorem}[section]
\newtheorem{cor}{Corollary}[section]
\usepackage{amsfonts}

\def\un{\hbox{$\mit I$\kern-.77em$\mit I$}}
\def\0{\hbox{$\mit I$\kern-.70em$\mit O$}}
\def\R{{\mathbb R}}
\def\T{{\mathbb T}}
\def\Z{{\mathbb Z}}
\def\N{{\mathbb N}}
\def\C{{\mathbb C}}
\def\P{\hbox{$\mit I$\kern-.277em$\mit P$}}

\begin{document}

\title{Spectral Analysis of Unitary Band Matrices}
\author{Olivier Bourget$^1$, James S. Howland$^2$, Alain Joye$^1$}
\date{ }
\maketitle
{\normalsize$^1$Institut Fourier, 
Universit\'e de Grenoble 1, BP 74, 38402 St.-Martin d'H\`eres,  
France\\
\indent
$^2$Department of Mathematics, University of Virginia, 
Charlottesville, VA 22903, USA}
\abstract{This paper is devoted to the spectral properties of 
a class of unitary operators with a matrix representation 
displaying a band structure. Such band matrices appear as 
monodromy operators in the study of certain quantum dynamical 
systems. These doubly infinite matrices essentially depend 
on an infinite sequence of phases which govern their spectral 
properties. We prove the spectrum is purely singular for random 
phases and purely absolutely continuous in case they provide 
the doubly infinite matrix with a periodic structure in the diagonal 
direction. We also study some properties of the singular 
spectrum of such matrices considered as infinite in one direction 
only.}

\section{Introduction}

The dynamical stability of quantum systems governed by a time
periodic Hamiltonian is often characterized in terms of the 
spectral properties of the corresponding monodromy operator,
a unitary operator defined as the evolution generated by the
Hamiltonian over a period. 
A first rough classification
consists in determining whether or not the spectrum of the monodromy
operator contains an absolutely continuous (a.c.) component.
The presence of absolutely continuous spectrum is a signature of 
unstable quantum systems, whereas 
a purely singular spectrum is a characteristic of quantum stability.

For smooth Hamiltonians, these spectral properties can be obtained 
through the study of an associated self-adjoint operator, the
so-called Floquet or quasi-energy operator \cite{h1}, \cite{y}. 
In case the Hamiltonian is singular, {\sl e.g.} when it corresponds 
to a kicked system, one is often lead to
consider the monodromy operator directly \cite{c2}. In both
situations, one is typically confronted with a problem where a
dense pure point operator is perturbed either by the addition
of a self-adjoint operator in the first case, or by a multiplicative 
unitary perturbation is the second case.
A more or less detailed spectral analysis can thus be performed 
provided a perturbative framework of some sort is available, or 
in case disorder is present. See {\sl e.g.}
\cite{be}, \cite{ds1}, \cite{dlsv}, \cite{gy}, \cite{h2}, \cite{h3},
\cite{n}, \cite{j} for the smooth case and, besides the review \cite{c2},
\cite{c1}, \cite{do}, \cite{ade}, \cite{bo} for the kicked case.

The dynamical quantum systems we address here are characterized 
by a monodromy operator given by a product of two pure point
unitaries, neither of which can be considered a perturbation of the other.
However, the spectral analysis can be carried over under certain
circumstances due to the fact the monodromy operator has a band 
structure in some basis. The motivation of the
construction of such operators is borrowed from the work \cite{bb}
which we briefly recall below.

As noted by these authors, this structure allows us to adapt the techniques 
developed in the study of one dimensional discrete Schr\"odinger 
operators to the unitary framework in order to obtain results about 
the spectrum of such monodromy operators. 

Let us briefly summarize the paper. In section 2, we define explicitly
the class of unitary operators on the integer lattices $\Z$ and $\N$ that
we shall study and discuss their relationship to \cite{bb}. These operators
depend on transmission and reflection amplitudes at each lattice point.
Some simple perturbative results for essential and absolutely continuous 
spectra are obtained in section 3. Here, the moduli of the transmission
and reflection amplitudes may vary from point to point, but in the remainder
of the paper these {moduli} are assumed constant on the lattice. In sections
4 and 5, we consider the {random} case, in which the phases are independent
and randomly distributed on the circle, and we prove that the spectrum is 
{purely singular}. To do this, we first establish a version of the Ishii-Pastur
Theorem according to which the absolutely continuous part of the spectrum is 
almost surely supported on the closure of the set where the Lyapunov exponent
vanishes and then prove that the Lyapunov exponent is everywhere positive. In section 6, we consider the {coherent} case, in which the phases are 
eventually periodic. We identify the absolutely continuous spectrum, and show
that the singular continuous spectrum is absent. Finally, in section 7, we give 
an example in which the phases are {almost periodic} and the spectrum is
{purely singular continuous}.

\setcounter{equation}{0}
\section{Construction of the Monodromy Operator}

We consider a
class of monodromy operators whose construction is motivated 
by the study of a model of electronic transport in a ring threaded 
by a linear time dependent magnetic flux, as discussed in \cite{bb},
and references therein. Neglecting the curvature of the ring, 
the instantaneous Hamiltonian of the one-body Schr\"odinger 
operator corresponds to that
of a one dimensional Schr\"odinger operator with a periodic potential
describing the material of the ring and time dependent boundary
conditions of Floquet type. With a choice of linear flux, the time 
plays the role
of the quasi-momentum. Therefore, as a function of time, the
Hamiltonian is periodic and its instantaneous spectrum is given by the band
structure corresponding to the potential. Under some adiabaticity
condition, the evolution operator is assumed to couple states
by adjacent pairs of states only by means of the Landau-Zener
mechanism. The concerned states are those whose corresponding eigenvalues 
become close to one another. Thus, a given state with index $k$ say, 
is coupled once
to the one with index $k-1$ and once with the one index
$k+1$. This yields the band structure of the evolution operator 
over a period in the basis of eigenvectors at time zero, say. 
We refer the reader to this paper for physical background 
and further description of the regime in which the model holds.
Let us now define our monodromy operator following the main
lines of the construction sketched above.

Our separable Hilbert space is $l^2(\Z)$ and we denote the canonical
basis by $\{ \ffi_k \}_{k\in \Z}$. In order to make contact with the
above model, we shall also state results for $l^2(\N)$.
The most general $2\times 2$ unitary matrix depends on $4$ parameters
and can be written as 
\be\label{gen22}
S=\E^{-i\theta}\pmatrix{r \E^{-i\alpha} & it \E^{i\gamma}\cr
                        it \E^{-i\gamma} & r \E^{i\alpha}},
\ee
where $\alpha,\gamma, \theta$ belong to the torus $\T$ and
the real parameters $t, r$, also called reflection and transition 
coefficients, are linked by $r^2+t^2=1$. 

We introduce an infinite set of such matrices $\{S_k\}_{k\in\Z}$
where $S_k$ depends on the phases $\alpha_k,\gamma_k, \theta_k$,
and the reflection and transition coefficients $t_k, r_k$. 
They are the building blocks of our monodromy operator in $l^2(\Z)$. 

Let $P_{j}$ be the orthogonal projector on the span of 
$\ffi_j, \ffi_{j+1}$ in $l^2(\Z)$.
We introduce $U_e, U_o$ two $2\times 2$ block diagonal unitary 
operators on $l^2(\Z)$
defined by
\bea\label{ueuo}
U_e &=& \sum_{k\in \Z} P_{2k}S_{2k}P_{2k}\nonumber\\
U_o &=& \sum_{k\in \Z} P_{2k+1}S_{2k+1}P_{2k+1},
\eea
or, in matrix representation in the canonical basis,
\be\label{22blocks}
U_e = \pmatrix{   \ddots & & & & \cr
                   &S_{-2}& & &   \cr
                   & &S_0& &  \cr
                   && &S_2& \cr    
                  & & & & \ddots     }
\ee
and similarly for $U_o$, with $S_{2k+1}$ in place of $S_{2k}$. 
Note that the $2\times 2$ blocks in  $U_e$ are shifted
by one with respect to those of $U_o$ along the diagonal. 

We now define the monodromy operator $U$, object of our
investigations, by 
\be\label{mono}
U=U_o U_e,
\ee
such that, for any $k\in\Z$,
\bea\label{matel}
U\ffi_{2k}&=&ir_{2k}t_{2k-1}\E^{-i(\theta_{2k}+\theta_{2k-1})}
\E^{-i(\alpha_{2k}-\gamma_{2k-1})}\ffi_{2k-1}\nonumber\\
&+&r_{2k}r_{2k-1}\E^{-i(\theta_{2k}+\theta_{2k-1})}
\E^{-i(\alpha_{2k}-\alpha_{2k-1})}\ffi_{2k}\nonumber\\
&+&ir_{2k+1}t_{2k}\E^{-i(\theta_{2k}+\theta_{2k+1})}
\E^{-i(\gamma_{2k}+\alpha_{2k+1})}\ffi_{2k+1}\nonumber\\
&-&t_{2k}t_{2k+1}\E^{-i(\theta_{2k}+\theta_{2k+1})}
\E^{-i(\gamma_{2k}+\gamma_{2k+1})}\ffi_{2k+2}\nonumber\\
& & \nonumber\\
U\ffi_{2k+1}&=&-t_{2k}t_{2k-1}\E^{-i(\theta_{2k}+\theta_{2k-1})}
\E^{i(\gamma_{2k}+\gamma_{2k-1})}\ffi_{2k-1}\nonumber\\
&+&it_{2k}r_{2k-1}\E^{-i(\theta_{2k}+\theta_{2k-1})}
\E^{i(\gamma_{2k}+\alpha_{2k-1})}\ffi_{2k}\nonumber\\
&+&r_{2k}r_{2k+1}\E^{-i(\theta_{2k}+\theta_{2k+1})}
\E^{i(\alpha_{2k}-\alpha_{2k+1})}\ffi_{2k+1}\nonumber\\
&+&ir_{2k}t_{2k+1}\E^{-i(\theta_{2k}+\theta_{2k+1})}
\E^{i(\alpha_{2k}-\gamma_{2k+1})}\ffi_{2k+2}.
\eea
In matrix form, without expliciting the elements, we have the
structure
\be\label{struct}
U=\pmatrix{\ddots & & & & & & & & &\cr
          \ast  &\ast &\ast &\ast & & & & & &\cr
          \ast  &\ast &\ast &\ast & & & & & &\cr
           & &\ast & \ast&\ast &\ast & & & &\cr
           & &\ast &\ast &\ast &\ast & & & &\cr
           & & & &\ast &\ast &\ast &\ast & &\cr
           & & & &\ast &\ast &\ast &\ast & &\cr 
           & & & & & &\ast &\ast &\ast &\ast \cr
           & & & & & &\ast &\ast &\ast&\ast  \cr
           & & & & & & & & &  \ddots           }   .
\ee

In the regime considered in \cite{bb}, 
the transition coefficients $t_k$, and all elements of the 
scattering matrices $S_k$ can be computed from 
the band functions of the periodic background potential. 
In particular, the transition coefficients may admit a limit as 
$k\ra\infty$.

In this paper, we briefly show how to get information on the 
spectral properties of the monodromy operators defined on
$l^2(\N)$ from those of operators defined on $l^2(\Z)$. Also,
we briefly demonstrate how spectral properties of $U$ when 
the $t_k$'s have limits $t_{\pm}$ as $k\ra\pm\infty$ 
can be related to those of the limiting operator with constant 
(in $k$) transition coefficients $t$.
Then we focus on the case of constant 
transition and reflexion coefficients
$t_k=t\in ]0,1[  \Leftrightarrow  r_k=r\in ]0,1[$, $\forall k\in \Z$,
which is the main object of our analysis.
This corresponds to a regime 
of the original model in which the sole behavior of the scattering phases 
$\theta_k, \gamma_k, \alpha_k$ determine the spectral properties of $U$.
It is argued in \cite{bb} on the basis of numerical computations that 
in case these phases have a coherent behavior
as functions of $k$, if they are periodic say, $U$ has an a.c. component 
in its spectrum, whereas $U$ should be singular if some phases are random. 
Following their arguments, we are aiming at a rigorous version of similar statements in our setting.

\setcounter{equation}{0}
\section{First properties}

At this point, we have slightly generalized the construction 
proposed by \cite{bb} in order to define our monodromy operator, 
a unitary pentadiagonal band matrix. Before going further in the analysis, 
one can ask whether
simpler unitary band matrices could provide interesting models,  
spectrally speaking,
as is the case in the self-adjoint setting where the discrete Schr\"odinger
operators are tridiagonal, though non-trivial. The next lemma answers
this question negatively, validating our model from another point of view.
Its proof can be found in Appendix.
\begin{lem}\label{trivtri}
If $U$ is unitary and tridiagonal,
then $U$ is either a shift operator, or an infinite direct sum of $2\times 2$ 
unitary matrices of the form (\ref{22blocks}). 
\end{lem}
On the other hand, it is straightforward to construct unitary band matrices
with larger width starting with general unitary finite size matrices, 
following the same steps as above.

\subsection*{Perturbative results}
In the physical context alluded to above, the natural Hilbert space is  
$l^2(\N^*)$, with $\N^*$ the set of positive integers, and the definition of the unitary monodromy operator, say $U^+$, is 
\bea\label{defup}
&&U^+\ffi_1=r_1e^{-i(\theta_0+\theta_1)}e^{-i\alpha_1}\ffi_1+
it_1e^{-i(\theta_0+\theta_1)}e^{-i\gamma_1}\ffi_2,\nonumber\\
&&U^+\ffi_k, \,\,  k>1, 
\eea 
as in (\ref{matel}). We shall also define $U^-$ on $l^2(-\N^*)$ in
a similar fashion. 
Consider $U^e$ on $l^2(\Z)$ defined by
(\ref{matel}) with even matrix elements
\be\label{defue}
\{t_{-k}, \theta_{-k},\alpha_{-k},\gamma_{-k}\}  = \{
t_{k}, \theta_{k}, \alpha_{k},\gamma_{k}\} \,\,\,\,
\forall k\in \N .
\ee
\begin{thm}\label{22} Let $U^+$ and $U^e$ be as above and 
let $U^+_{a.c.}$ and $U^e_{a.c.}$ denote their restriction to their 
respective absolutely continuous subspaces. Then
$$\sigma_{{ess}}(U^+)=\sigma_{{ess}}(U^e), \,\,\,\,
\mbox{ and }\,\, U^+_{a.c.}\oplus U^+_{a.c.}\simeq U^e_{a.c.},$$
where $\simeq$ means unitary equivalence.
\end{thm}
{\bf Proof:} 
We can write on $l^2(-\N^*)\oplus \C \oplus l^2(\N^*)$
\bea
U^e&=& \pmatrix{C U^+ C^{-1} & &
\cr &1& \cr
& & U^+}  +F\nonumber\\
&=&\pmatrix{ C & &
\cr &1& \cr
& & \un}\pmatrix{U^+  & &
\cr &1& \cr
& & U^+} \pmatrix{ C^{-1} & &
\cr &1& \cr
& & \un} +F
\eea 
where absent elements denote zeros, $\un$ is the identity, $C$ is the operator 
\be
\matrix{C:&l^2(\N^*)\ra l^2(-\N^*)\cr
              & \ffi_k\mapsto\ffi_{-k}}
\ee
and $F$ is a finite rank operator.
Noting that $\sigma(C U^+ C^{-1})=\sigma( U^+)$, we get 
the result by Weyl's and Birman-Krein's theorems on invariance
of essential, resp. absolutely continuous spectrum, under
compact, resp. trace class, perturbation. \ep

Let us now consider the situation where the transition coefficients 
of the operator $U$ defined by (\ref{matel}) satisfy
\be\label{limt}
\lim_{k\ra\pm\infty}t_k=t_{\pm}\Longleftrightarrow 
\lim_{k\ra\pm\infty}r_k=r_{\pm}.
\ee
We measure the convergence by means of the quantities 
$\delta^{\pm}$ defined by
\be
\delta^+(j)=\max\{r_jr_{j-1}-r_{+}^2\, ,\,\,
t_jt_{j-1}-t_{+}^2\, ,\,\, t_jr_{j\pm 1}-r_{+}t_{+} \}\, ,\,\, j\in\N,
\ee
and similarly for $\delta^{-}$.
Let $U^{\pm}(t_{\pm})$ defined on $l^2(\pm \N^*)$ by (\ref{defup}) with $t_k=t_{\pm}$ and $r_k=r_{\pm}$, 
for all $k\in\pm\N^*$ .
\begin{thm}\label{32} Assume (\ref{limt}) and let $U$ and $U^{\pm}(t_{\pm})$ be as above. 
Then
$$\sigma_{{ess}}(U)=\sigma_{{ess}}(U^+(t_+))\cup\sigma_{{ess}}(U^-(t_-)). $$
If, furthermore, there exists $\epsilon>1/2$ such that 
$\sup_{j\in\N}\delta^{\pm}(j)j^{2\epsilon}<\infty$,
$$U_{{a.c.}}\simeq U^+_{{a.c.}}(t_+)\oplus U^-_{{a.c.}}(t_-). $$
\end{thm}
{\bf Proof:} 
Let us introduce the asymptotic unitary operator $U_{-,+}$ by
\be\label{asop}
U_{-,+}=\pmatrix{U^-(t_-)  & &
\cr &1& \cr
& & U^+(t_+)}.
\ee
The difference between the actual and asymptotic operators 
is given by the operator 
\be
\Delta=U-U_{-,+}
\ee
 whose matrix elements
$\Delta(j,k)=\bra\ffi_j | \Delta \ffi_k\ket $ satisfy for $|k|>1$,
\be\label{madel}
|\Delta(j,k)|\leq\left\{\matrix{\max\{\delta^{\pm}(j-1),\delta^{\pm}(j),
\delta^{\pm}(j+1)\}& \mbox{ if } |j-k|\leq 2\cr
 0 &  \mbox{ otherwise. } }\right.
\ee
Therefore, approximating $\Delta$ by a finite matrix $\Delta_N$, 
we can use the Schur condition, \cite{k}, P.143, to estimate the norm of
the difference $ \Delta-\Delta_N$ and get $\|\Delta-\Delta_N\|\ra 0 $ 
as $N\ra\infty$. This, in turn, shows that 
$\Delta$ is compact and that the essential spectra of $U$ and $U_{-,+}$
coincide and yields the first assertion.
The second is proven following arguments used in \cite{h2}.
Let $\epsilon>1/2$ and set $<j>=(1+j^2)^{1/2}$. We define 
$\Lambda=\mbox{diag }\{<j>^{\epsilon}\}$ in the
basis $\{\ffi_k\}_{k \in\Z}$. As $ \Delta=\Lambda^{-1}(\Lambda \Delta
\Lambda)\Lambda^{-1}$, where $\Lambda^{-1}$ is Hilbert-Schmidt, 
$\Delta$ will be trace class as soon as $\Lambda \Delta\Lambda$ is bounded.
Its non-zero matrix elements are 
\be
(\Lambda \Delta\Lambda)(j,k)=(<j><k>)^{\epsilon}\Delta(j,k), \,\,\, k=j, 
j\pm 1,j\pm 2,
\ee
so that we get boundedness as above from the Schur condition and
the estimate (\ref{madel}).\ep\\
{\bf Remarks:} \\
i) An analogous statement is obviously true for operators defined on
$l^2(\N^*)$.\\
ii) The condition $\sup_{j\in\N}\delta^{\pm}(j)j^{2\epsilon}
<\infty$ for some $\epsilon>1/2$, actually is necessary as well to have $\Delta$ trace-class.
Indeed, in case $t_{\pm}=t \Longleftrightarrow r_{\pm}=r$ with $tr\neq 0$,
$t\neq r$, and $t-t_j=1/j^{\alpha}$, one checks that $\delta(j)\sim 
c/j^{\alpha}$, for some constant $c$. Assuming that 
$\Delta$ is trace class, we have $\sum_{j\in\Z}|\Delta(j,j)|<\infty$
which is equivalent to $\sum_{j\in\Z}1/<j>^{\alpha}<\infty$ and requires
$\alpha>2\epsilon$, for some $\epsilon>1/2$.\\
iii) It is clear that similar perturbative results holds for
more general cases where the phases have a limiting behavior as well.

The case $t_{+}=t_-=0$ is of particular interest and allows a stronger
result. 
\begin{thm}\label{ppp} Consider $U$ on $l^2(\Z)$ defined by (\ref{matel}) and $U^+$ on 
$l^2(\N^*)$ defined by (\ref{defup}). 
If $\liminf_{k\ra\pm\infty} {t_k}=0$, then
\be
\sigma_{a.c.}(U)= \sigma_{a.c.}(U^+)=\emptyset.
\ee
\end{thm}
{\bf Proof:} We consider $U$ only, the proof for $U^+$ being similar. 
Let $U_n$ be equal to $U$ with $t_n=0$ and  
\be
F_n=U-U_n .
\ee 
The matrix of $U_n$ is separated into two  
disjoint blocks and $F_n$ is a rank four operator with 
$\|F_n\|\leq c t_n$. The hypothesis insures the existence 
of a subsequence $t_{n(k)}$ going to zero as fast as we wish,
say as $<k>^{-2}$, when $k\ra\pm\infty$. We set 
\be
G=\sum_{k\in\Z}F_{n(k)}\,\, \mbox{ and } \,\, \tilde{U}=U-G.
\ee 
By construction, we have for some constant $\tilde{c}$
\be
\|G\|_1\leq 4\sum_{k\in\Z}\|F_{n(k)}\|\leq
 \sum_{k\in\Z}\frac{\tilde{c}}{<k>^2}<\infty
\ee
and $\tilde{U}$ is pure point, hence the result. \ep \\
{\bf Remarks:} \\
i) In case there exists a subsequence $\{t_{n(k)}\}$ such that
\be
\lim_{k\ra +\infty} t_{n(k)}=t_+ \,\,\mbox{ and }\,\,\lim_{k\ra -\infty} t_{n(k)}=t_-,
\ee
a similar construction is valid and we get an approximation of the form
\be
U=\tilde{U}_{-,+}+G_{-,+},
\ee
where $\tilde{U}_{-,+}$ contains an infinite number of $t_-$ and $t_+$ in its matrix representation and $ G_{-,+}$ is trace class. However we do not know the spectral properties of such $\tilde{U}_{-,+}$'s.\\
ii) If $U^+(0)$ defined according as in Proposition (\ref{ppp}) is such that
its pure point spectrum possesses a finite number of accumulation 
points only, then, if $\lim t_j\ra 0$, $\sigma(U^+)$ is pure point with 
finitely many accumulation points as well. This will be true in case the
phases have a coherent behavior, see Section \ref{coher}.

Motivated by the previous theorems, we now address the spectral properties
 of the limiting operators.

\subsection*{Constant reflection and transition 
coefficients}

From now, $t_k=t$, $r_k=r$, $\forall k\in\Z$. 
We first note that the extreme cases where $rt=0$ are spectrally trivial.
\begin{prop}\label{triv}
In case $t=0\Leftrightarrow r=1$, $U$ is pure point and
if  $t=1\Leftrightarrow r=0$, $U$ is purely absolutely continuous.
\end{prop} 
{\bf Proof:} The first case is trivial. In the second case, we observe
that $U$ is reduced by the supplementary subspaces $L^+$, 
respectively $L^-$, generated by the vectors in the canonical 
basis with even indices, respectively odd indices. Moreover 
$U|_{L^{\pm}}$ is unitarily equivalent to the shift operator, hence
the result.  \ep\\
{\bf Remark:} As a typical Corollary, we get the following spectral 
properties for monodromy operators $U^+$ defined on $l^2(\N^*)$ according 
to (\ref{defup}).
\be
\sigma_{a.c.}(U^+)=S^1 \mbox{ if } t_j-1\sim 1/j^{\alpha}, \,\,\,  
\alpha>1.
\ee

The rest of the paper is devoted to studying
the limiting operator when  
$t_k=t\in ]0,1[  \Leftrightarrow  r_k=r\in ]0,1[$, $\forall k\in \Z$.

All phases in the definition of $U$ do not play the same role, as
the following Lemma shows. One the one hand it justifies the choice 
made in \cite{bb} where the phases $\gamma_k$ are taken equal to zero and, 
on the other hand, it will be very useful below.
\begin{lem}\label{free}
If we denote the matrix (\ref{struct})  by $M(\{ \theta_k\},\{ \alpha_k\},\{\gamma_k \})$,
and identify $U$ to it, we have for any sequences $\{ \theta_k\},\{ \alpha_k\},\{\gamma_k \}$, $k\in\Z$
$$
U\equiv M(\{ \theta_k\},\{ \alpha_k\},\{\gamma_k \})
\simeq  M(\{ \theta_k\},\{ \alpha_k\},\{0 \}).
$$
\end{lem} 
{\bf Remarks:}\\
i) As a Corollary, we can replace the sequence $\{\gamma_k\}$, $k\in\Z$ in the 
definition of $U$ by any other sequence $\{\gamma_k'\}$.\\
ii) The same statement is true for $U^+$ defined on $l^2(\N^*)$ by 
(\ref{defup}).\\
{\bf Proof:}
Let $V$ be the unitary operator defined by
\be\label{unit}
 V\ffi_k=\E^{i\zeta_k}\ffi_k,\,\,\,  k\in\Z .
\ee
One checks easily that the operator $V^{-1}UV$ has the form
$M(\{ \theta_k\},\{ \alpha_k\},\{0 \})$ in the canonical basis 
provided for all $j\in\Z$,
\be
\zeta_j-\zeta_{j-1}= -\gamma_{j-1}.
\ee
This is realized by taking, for example, $\zeta_0=0$ and
\be
\zeta_k=-\sum_{j=0}^{k-1}\gamma_j, \,\,\,\,\, \zeta_{-k}=
\sum_{j=-1}^{-k}\gamma_j, \,\,\,\,\,  k\in \N^*.
\ee
 \hfill \ep

\subsection*{Generalized eigenvectors}
Without making use yet of the freedom we have in the sequence
$\{\gamma_k\}$, $k\in \Z$, we now turn  to the eigenvalue equation 
\bea\label{eveq}
&&U\psi=\E^{i\lambda}\psi, \nonumber \\  
&&\psi=\sum_{k\in\Z}c_k\ffi_k, \,\, c_k\in \C, \,\, \lambda\in\C.
\eea
One sees from the structure (\ref{struct}) of the operator $U$, that
if $\psi$ satisfies (\ref{eveq}),  
a linear relation between the coefficients $(c_{2k}, c_{2k+1})$ and
$(c_{2k-2}, c_{2k-1})$ of the form 
\be\label{trans}
\pmatrix{c_{2k} \cr c_{2k+1}}=T(k)\pmatrix{c_{2k-2} \cr c_{2k-1}}
\ee
must exist, provided some $2\times 2$ matrix is invertible. Using the 
definition (\ref{matel}), straightforward computations show that the matrix
$T(k)$ has elements 
\bea\label{deft}
T(k)_{11}&=& -\E^{-i(\lambda+\gamma_{2k-1}+
\gamma_{2k-2}+\theta_{2k-1}+\theta_{2k-2})} \\
T(k)_{12}&=&i\frac{r}{t}\left(
\E^{-i(\lambda+\gamma_{2k-1}-
\alpha_{2k-2}+\theta_{2k-1}+\theta_{2k-2})}
-
\E^{-i(\gamma_{2k-1}-\alpha_{2k-1})}\right)\nonumber\\
T(k)_{21}&=&i\frac{r}{t}\left(\E^{-i(\theta_{2k-2}-\theta_{2k}+
\gamma_{2k}+\gamma_{2k-1}+\gamma_{2k-2}+\alpha_{2k-1})}\right.\nonumber\\
&-&\left.\E^{-i(\lambda+\theta_{2k-2}+\theta_{2k-1}+
\gamma_{2k}+\gamma_{2k-1}+\gamma_{2k-2}+\alpha_{2k})}\right)\nonumber\\
T(k)_{22}&=&-\frac{1}{t^2}\E^{i(\lambda+\theta_{2k}+\theta_{2k-1}-
\gamma_{2k}-\gamma_{2k-1})}\nonumber\\
&+&\frac{r^2}{t^2}\E^{-i(\gamma_{2k}+\gamma_{2k-1})}\left(
\E^{i(\theta_{2k}-\theta_{2k-2}+\alpha_{2k-2}-\alpha_{2k-1})}
+\E^{-i(\alpha_{2k}-\alpha_{2k-1})}\right)\nonumber\\
&-&\frac{r^2}{t^2}\E^{-i(\lambda+\theta_{2k-2}+\theta_{2k-1}+
\gamma_{2k}+\gamma_{2k-1}+\alpha_{2k}-\alpha_{2k-2})},\nonumber
\eea
provided $t\neq 0$.
We also compute 
\be\label{det=1}
  \det T(k)=\E^{-i(\theta_{2k-2}-
\theta_{2k}+\gamma_{2k}+
2\gamma_{2k-1}+\gamma_{2k-2})}
\ee
so that $| \det T(k)|=1$.

Therefore, once the coefficients $(c_0, c_1)$ are given, we compute
for any $k\in\N$,
\bea\label{cocycle}
\pmatrix{c_{2k} \cr c_{2k+1}}&=&T(k)\cdots T(2)T(1)\pmatrix{c_{0} \cr c_{1}}
 \equiv \Phi(k)\pmatrix{c_{0} \cr c_{1}}\nonumber \\
\pmatrix{c_{-2k} \cr c_{-2k+1}}&=&T(-k+1)^{-1}\cdots 
T(-1)^{-1}T(0)^{-1}\pmatrix{c_{0} \cr c_{1}}
\equiv \Phi(-k)\pmatrix{c_{0} \cr c_{1}}.
\eea
The multiplicity of possible eigenvalues is therefore bounded
by two.

\setcounter{equation}{0}
\section{Random setting}

Apart from the fact 
that our transfer matrix is complex valued instead of the usual real valued 
setting suiting the discrete Schr\"odinger case, we will see
that here also one Lyapunov exponent is enough to describe the 
spectral properties of $U$, when the phases are random and the transfer 
matrices $T(k)$ are independent and identically distributed.

Making use of Lemma \ref{free}, let us introduce a probabilistic space 
$(\Omega, {\cal F}, \P)$, where $\Omega$ is identified with $\{{\T}^{\Z} \}$,
$\T$ being the torus,
and $\P=\otimes_{k\in\Z}\P_0$, where $\P_0$ is the uniform distribution on $\T$
with ${\cal F}$ the $\sigma$-algebra generated by the cylinders. We introduce
the set of random vectors on 
$(\Omega, {\cal F}, \P)$ given by 
\bea\label{beta}
& &\beta_k=(\theta_k, \alpha_k): \Omega \rightarrow \T^2,
\,\,\, k\in \Z,\nonumber\\
& &\theta_k(\omega)=\omega_{2k}, \,\,\,\, \alpha_k(\omega)=\omega_{2k+1}.
\eea
The random vectors $\{\beta_k\}_{k\in\Z}$ are
i.i.d and uniformly distributed on $\T^2$.

We denote by $U_{\omega}$
the random unitary operator corresponding to the random
infinite matrix (\ref{struct})
\be\label{ranun}
U_{\omega}=M(\{ \theta_k(\omega)\},\{ \alpha_k(\omega)\},\{0\}).
\ee

Introducing the shift operator $S$ on $\Omega$ by
\be
  S(\omega)_k=\omega_{k+2}, k\in\Z, 
\ee
we get an ergodic set $\{S^j\}_{j\in\Z}$ of translations.
With the unitary operator $V_j$ defined on the canonical basis of 
$l^2(\Z)$ by
\be
V_j\ffi_k=\ffi_{k-2j}, \forall k\in\Z,
\ee
we observe that for any $j\in\Z$
\be\label{ero}
U_{S^j\omega}=V_jU_{\omega}V_j^*.
\ee
Therefore, our random operator $U_{\omega}$ is a an ergodic operator. 
The spectral projectors $E_{\Delta}(\omega)$ of $U_{\omega}$ 
where $\Delta$ is a Borel set of $\T$, define a weakly 
measurable projector valued family of operators on $\Omega$ and the
spectrum of $U_{\omega}$ is deterministic, see \cite{cl}.
However, we shall not make use of these properties below.

As it stands, the transfer matrix $T(k)$ depending on the random vectors
$\beta_{2k}, \beta_{2k-1}, \beta_{2k-2}$ seems to be correlated with 
$T(k+1)$ and $T(k-1)$. 
Using the same Lemma \ref{free}, we can replace the sequence 
$\{0\}$ in (\ref{ranun}) by $\{(-1)^{k+1}\alpha_k \}$, so that we 
consider explicitly
$M(\{ \theta_k\},\{ \alpha_k\},\{(-1)^{k+1}\alpha_k \})$ and
the corresponding transfer matrices.
Thus, in terms of the new variable, with $\lambda\in\R$,
\bea\label{nrv}
\eta_k(\lambda)=\theta_k+\theta_{k-1}+\alpha_k-\alpha_{k-1}+\lambda,
\eea
the transfer matrix can be written as
\be\label{mapt}
 T(k) \equiv T(\eta_{2k}(\lambda), \eta_{2k-1}(\lambda))\nonumber\\
\ee
with
\bea\label{tren}
T(k)_{11}&=& -\E^{-i\eta_{2k-1}} \\
T(k)_{12}&=&i\frac{r}{t}\left(\E^{-i\eta_{2k-1}(\lambda)}
-1\right)\nonumber\\
T(k)_{21}&=&i\frac{r}{t}\left(\E^{i(\eta_{2k}(\lambda)-\eta_{2k-1}(\lambda))}
-\E^{-i\eta_{2k-1}(\lambda)}\right)\nonumber\\
T(k)_{22}&=&-\frac{1}{t^2}\,\E^{i\eta_{2k}(\lambda)}
+\frac{r^2}{t^2}\left(\E^{i(\eta_{2k}(\lambda)-\eta_{2k-1}(\lambda))}
+1-\E^{-i\eta_{2k-1}(\lambda)}\right).\nonumber
\eea

Therefore, introducing the set of random vectors 
\be\label{defdel}
\delta_{k}=(\eta_{2k}(\lambda),\eta_{2k-1}(\lambda) ) \in \T^2,\,\, k\in\Z
\ee 
we observe that the set of random transfer matrices $\{T(k)\}_{k\in\Z}$ will
be independent provided the set of random vectors 
$\{\delta_{k}\}_{k\in\Z}$ are independent. 

Using properties of the characteristic functions of random vectors 
\be
\Phi_{\beta}(n_1, n_2)={\mathbb E}(e^{-i(n_1\beta_1(\omega)+ n_2\beta_2(\omega))}), \,\,\, n_1, n_2 \in \Z,
\ee 
we get the following Lemma.
\begin{lem}\label{indep} If the vector $\{\beta_k\}_{k\in\Z}$ are i.i.d and uniform,
the random vectors 
$\{\delta_{k}\}_{k\in\Z}$ are also i.i.d and uniformly 
distributed on $\T^2$. In turn, the set 
of transfer matrices $\{T(k)\}_{k\in\Z}$ are i.i.d random 
matrices in $Gl_2(\C)$. 
\end{lem}

We can now state our main result in the random setting:
\begin{thm}\label{main}
Let $U_{\omega}$ be defined by its matrix elements (\ref{matel})
with $t\in (0,1)$. Assuming the phases $\{\alpha_k\}_{k\in\Z}$ 
and $\{\theta_k\}_{k\in\Z}$ are i.i.d. and uniform on $\T$, 
we have almost surely
$$ \sigma_{a.c.}(U_{\omega})=\emptyset .$$
\end{thm}

The next section is devoted to the proof of this Theorem.\\

\noindent{\bf Remarks:} \\
i) The same result for $U^+_{\omega}$ defined by (\ref{defup}) holds
by Theorem \ref{32}.\\
ii) In case the phases $\alpha_k\in \T$ are deterministic
and of the form $\alpha_k=a k +b$,  $a, b\in \R$, whereas the 
$\theta_k$'s are i.i.d. and uniform, the conclusions of the above Lemma
and Theorem still hold. The same is true if the $\theta_k$'s are 
deterministic and constant whereas the $\alpha_k$'s are i.i.d. 
and uniform.\\
iii) To motivate our hypotheses on the uniform distribution of the phase vectors 
$\beta_k$, we  recall the 
\begin{lem}\label{shortproof}
If $X_k$, $k\in \Z$, is a set of i.i.d. 
random variables on $\T$ with support not reduced to a point, then 
the random variables $Y_k^{\pm}=X_k \pm X_{k-1}$, $k\in\Z$ are
independent if and only if the $X_k$ are uniformly distributed.
\end{lem}
A proof of Lemmas \ref{indep} and \ref{shortproof}
can be found in Appendix.

\setcounter{equation}{0}
\section{Lyapunov Exponents}

As the map $(\ref{nrv})$ is measurable, 
we can realize our transfer matrices as an i.i.d. random process 
on the same probabilistic space $(\Omega, {\cal F}, \P)$ in such a way
that
\be
T(k,\omega)=T(\omega_{2k},\omega_{2k-1}), k\in\Z, \,\, 
\forall \omega\in \T^{\Z},
\ee
with the $C^{\infty}$ map $T: \T^2 \rightarrow Gl_2(\C)$ defined in
(\ref{mapt}).
Therefore,
\be
  T(k+1,\omega)=T(k, S(\omega)), \,\,\forall k\in\Z .
\ee
The set of translations $\{S^j\}_{j\in\Z}$ is ergodic
and we can write for all $k\in\N^*$,
\bea
\Phi(k,\omega)&=&T(k,\omega)T(k-1, \omega) \cdots T(1,\omega)
\nonumber\\
&=&T(1,S^k(\omega))T(1,S^{k-1}(\omega))\cdots T(1,\omega).
\eea
Similarly,
\be
  \Phi(-k,\omega)=T^{-1}(0,S^{-k+1}\omega)T^{-1}(0, S^{-k}\omega)\cdots 
T^{-1}(0,\omega).
\ee
Therefore $\{\Phi(k,\omega)\}_{k\in\N}$ defines a random ergodic linear 
dynamical system over $Gl_2(\C)$ generated by the map $T(1,\cdot)$
and $\{\Phi(-k,\omega)\}_{k\in\N}$ defines another one generated by 
$T^{-1}(0, \cdot)$. 

We are now formally in good shape to apply Oseledec's and Furstenberg's 
Theorems to define and study the Lyapunov exponents.
However, the last result is stated for real valued matrices, and, 
in particular, irreducibility properties of groups of matrices are 
a delicate matter. 
Therefore, we first want to map our problem to
a problem involving matrices in $Gl_4(\R)$. This is done very conveniently
using the method described in \cite{mt}, which we apply to our setting.
We will denote by  $\bra \cdot | \cdot  \ket$ the scalar product on
${\mathbb R}^4$ or  ${\mathbb C}^2$ and we introduce
\be
I = \left( \begin{array}{cc} 1 & 0 \\ 0 & 1 \end{array}
\right) \enspace , \enspace J = \left( \begin{array}{cc} 0 & 1 \\ -1 & 0 \end{array}
\right)\enspace .
\ee
We define a sub-algebra of ${\cal A}_4({\mathbb R})$ of ${\cal M}_4({\mathbb R})$
by
\be
{\cal A}_4({\mathbb R}) = \left\{ \left( \begin{array}{cc} a_1 I + a_2
        J & b_1 I + b_2J \\ c_1 I + c_2
        J & d_1 I + d_2J \end{array} \right) ,
        a_j, b_j, c_j, d_j \in {\mathbb R}, j=1,2. \right\}.
\ee
The topology on  ${\cal M}_2({\mathbb C})$, ${\cal M}_4({\mathbb R})$
is generated by the spectral norm 
\be
\| A \| = \sqrt{ \sum_{\lambda \in \sigma(|A|)} |\lambda|^2 } 
\ee
and that of ${\cal  A}_4({\mathbb R})$ 
is the induced topology.
Let $\rho$ be the mapping $ \rho:{\mathbb C}^2 \rightarrow {\mathbb R}^4$
\be
\pmatrix{ x\cr y } \rightarrow  \pmatrix{\Re(x) \cr -\Im(x)\cr \Re(y)\cr -\Im(y) } ,
\ee
and  $\tau:  {\cal M}_2({\mathbb C}) \rightarrow {\cal A}_4({\mathbb R})$ 
be defined by 
\be
 \pmatrix{a & b \cr c & d }\rightarrow 
  \pmatrix{\Re(a)I + \Im(a)J & \Re(b)I +
    \Im(b)J \cr \Re(c)I + \Im(c)J & \Re(d)I + \Im(d)J }.
\ee
The following properties are readily checked:
\begin{lem}\label{morphisme}
  For any $u, v \in {\mathbb
  C}^2$, and any $ \alpha \in {\mathbb C}$,
\bea
&&\rho(u+v) = \rho(u) + \rho(v) \\
&&\rho(\alpha u) = \Re(\alpha) \rho(u) + \Im(\alpha) \rho(i u) .
\eea
For any $ A, B \in {\cal
  M}_2({\mathbb C})$, and $ \alpha \in {\mathbb R}$,
\bea
&&\tau(A+B) = \tau(A) + \tau(B), \,\,\,\, \tau(A B) = \tau(A) \tau(B) ,\nonumber\\
&&\tau(\alpha A) = \alpha \tau(A), \,\,\,\,\hspace{1.6cm} \tau(A^*) = \tau(A)^* , 
\nonumber\\
&&\tau(A^{-1})= \tau(A)^{-1}.
\eea
The last formula meaning that if $A \in {\cal M}_2({\mathbb C})$ 
is invertible, $\tau(A)$ is also invertible and the formula is true.
Finally, for all $ u \in {\mathbb C}^2, \forall T \in {\cal M}_2({\mathbb
  C}), $
\be\label{413}
\rho (Tu) = \tau(T) \rho(u).
\ee
\end{lem}

We also note the following Lemma for future reference. 
\begin{lem}\label{vpdoubles}
If $A \in {\cal M}_2({\mathbb C})$ and $|\det(A)|=1$, then
$|\det(\tau(A))|=1$.
If $A$ is self adjoint with
eigenvalues $\gamma_1$ and $\gamma_2$ , then $\tau(A)$ is real symmetric
with eigenvalues $\gamma_1$ and $\gamma_2$ of multiplicity two.
\end{lem}

\noindent More general results of the same sort in higher dimension can be 
found in \cite{mt}.

\noindent{\bf Remarks:}\\
i) Let us note as a consequence of Lemma \ref{vpdoubles} that the mappings 
$\rho$ and $\tau$ are 
homeomorphisms and 
$\forall u \in {\mathbb C}^2$, $\forall A \in {\cal M}_2({\mathbb C})$,
\be\label{continuite}
\|\rho(u)\| = \|u\| ,\,\,\,
\|\tau(A)\| = \sqrt{2}  \| A\| .
\ee
ii) The mapping $\rho$ doesn't transport scalar product but 
it does preserve the norm. Note that we have for all  
$\forall u,v  \in {\mathbb C}^2$, and all $T
\in {\cal M}_2({\mathbb C})$, 
\bea
& &\bra \rho(iu)|\rho(u) \ket = 0\\
& &\bra \rho(u)|\tau(T)\rho(v) \ket = \bra \rho(iu)|\tau(T)\rho(iv) \ket
= \Re(\bra u|Tv \ket) ,\\
& &\bra \rho(i u)|\tau(T)\rho(v) \ket = -\bra \rho(u)|\tau(T)\rho(iv) \ket = 
\Im(\bra u|Tv \ket) .
\eea
Therefore, if $u$ and  $v$ are orthogonal in  ${\mathbb
  C}^2$, $\rho(u)$ and $\rho(v)$ are also orthogonal.

\subsection*{Existence of the Lyapunov Exponents}

Using this operator $\tau: Gl_2(\C)\rightarrow Gl_4(\R)$, 
we can now consider the random ergodic linear dynamical system over $Gl_4(\R)$ 
defined from $\{\Phi(k,\omega)\}_{k\in\N}$ by
\be\label{rlds}
\Psi(k,\omega)=\tau(\Phi(k,\omega))
\ee
generated by the map $\tau(T(1,\cdot)):\Omega \rightarrow  Gl_4(\R)$. 
We will work similarly if $-k\in\N$.

We now apply Oseledec's Theorem according to \cite{a}, 
Thm. 3.4.11, specialized to our setting.
 
\begin{prop}
Let the random ergodic dynamical system generated by the
map $\tau(T(1,\cdot )): \Omega\rightarrow Gl_4(\R)$. Then, 
on an invariant set $\Omega_0\subset\Omega$ of $\P$-measure
one, the following limit exists
\be
\lim_{n\ra\infty}(\Psi(n,\omega)^*\Psi(n,\omega))^{1/2n}=\Lambda(\omega).
\ee
The matrix $\Lambda(\omega)$ possesses at most $2$ distinct eigenvalues
of multiplicities $2$, denoted by
\be
\E^{\gamma_1}\geq \E^{\gamma_2}\equiv \E^{-\gamma_1} >0,
\ee
associated with at most two eigenspaces 
${\cal E}_1(\omega),{\cal E}_2(\omega)$.
The Lyapunov exponents $ \gamma_1\geq \gamma_2 $ are constant 
almost surely.\\
If $\gamma_1 > 0$, there exists a filtration of $\R^4$,  \\
$\{0\} \subset
  {\cal V} (\omega) \subset {\mathbb R}^4
$ such that
\be
{\cal V}(\omega) = {\cal E}_2(\omega) ,\,\,\mbox{ and }
\R^4={\cal E}_2(\omega)\oplus{\cal E}_1(\omega)
\ee
and  $u \in {\cal V}(\omega)$ iff
\be
\lim_{n \rightarrow +\infty} \frac{1}{n} \log \|\Psi(n,\omega) u \| = 
\gamma_2=-\gamma_1<0
\ee
and  $u \in \R^4\setminus {\cal V}(\omega)$
\be
\lim_{n \rightarrow +\infty} \frac{1}{n} \log \|\Psi(n,\omega) u \| = 
\gamma_1>0.
\ee
Moreover, there exists a splitting 
\be
\R^4=E_2(\omega)\oplus E_1(\omega)
\ee
such that
\be
\lim_{n \rightarrow \pm\infty} \frac{1}{n} \log \|\Psi(n,\omega) u \| =
\gamma_j \Leftrightarrow u\in E_j\setminus \{0\}
\ee
\end{prop}
{\bf Proof:}
We need to check the hypotheses of the Ergodic Multiplicative (Oseledec's) 
Theorem, see {\sl e.g.} \cite{a}, Thm. 3.4.11, in order to get the existence of the limit. All norms being equivalent, considering the maximum modulus of the 
matrix elements, we get the existence of a finite constant depending 
only on $0<t<1$ such that $C(t)^{-1}\leq \| T(1,\omega)\|\leq C(t)$. As 
$|\det (T(1,\omega))|=1$, the same bound is true for $T(1,\omega)^{-1}$.
The properties of $\tau$ finally yield
\be
  (\ln^+ \|\tau(T(1,\cdot ))\|+ \ln^+ \|\tau(T(1,\cdot )^{-1})\|)
\in L^1(\Omega, {\cal F}, \P),
\ee
where $ \ln^+ (x)= \max \{\ln (x), 0\}, x>0$, which ensures the existence 
of the limit. The statements about 
the number of Lyapunov exponents, their relations and multiplicities
are shown as follows. 
For any $n$, the $2\times 2$ matrix $\Phi(n,\omega)^*\Phi(n,\omega)$ is 
positive, of determinant one so that it either possesses two distinct 
eigenvalues $\sigma_1(n,\omega)>\sigma_2(n,\omega)
=1/\sigma_1(n,\omega)>0$ (of multiplicity one), 
or it is the identity matrix. Therefore,
$\Psi(n,\omega)^*\Psi(n,\omega)=\tau(\Phi(n,\omega)^*\Phi(n,\omega))$
has two distinct eigenvalues $\sigma_1(n,\omega)>\sigma_2(n,\omega)
=1/\sigma_1(n,\omega)>0$ of multiplicity two, 
or it is the identity matrix in $\R^4$. The determinant being continuous,
the limit $\Lambda(\omega)$ is also positive of determinant equal to one. 
By continuity of $\tau$ and $ \tau^{-1}$, $\Lambda(\omega)$ also belongs 
to ${\cal A}(\R)$ and there exists $\kappa(\omega)\in {\cal M}_2(\C)$ such that
$\tau(\kappa(\omega))=\Lambda(\omega)$. Moreover, the relation  
$\kappa(\omega)=\lim_{n\ra\infty}\Phi(n,\omega)^*\Phi(n,\omega)$ shows
that $\kappa(\omega)$ is also positive of determinant one, which proves
that the multiplicities of the eigenvalues of $\Lambda(\omega)$ is two
or it is the identity matrix. \ep

\begin{cor}\label{expdecr} Under the same hypotheses as above, 
there exists almost surely a subspace ${\cal  V}^0$ of $\C^2$ of complex dimension $1$  such that 
\bea
& & \forall u \in {\cal V}^0 \setminus \{0\}, \enspace
\lim_{n \rightarrow +\infty} \frac{1}{n} \ln \| \Phi(n,\omega) u \| 
= - \gamma_1 < 0\nonumber\\
& & \forall u \in {\mathbb C}^2 \setminus {\cal V}^0, \enspace
\lim_{n \rightarrow +\infty} \frac{1}{n} \ln \|\Phi(n,\omega) u \| = 
\gamma_1 > 0.
\eea
Also, there exists a splitting $\C^2=E_2^0(\omega)\oplus E_1^0(\omega)$
such that 
\be
\lim_{n \rightarrow \pm\infty} \frac{1}{n} \log \|\Phi(n,\omega) u \| =
\gamma_j \Leftrightarrow u\in E_j^0\setminus \{0\}.
\ee

\end{cor}
\noindent {\bf Proof: } By the Proposition, there exists a filtration: 
$\{0\} \subset {\cal V} \subset {\mathbb R}^4$,
such that:
\bea
& &\forall v \in {\mathbb R}^4 \setminus {\cal V} \enspace,
\lim_{n \rightarrow +\infty} \frac{1}{n} \log \|\tau(T(1,{S^{n} \omega}))
\ldots \tau(T(1, \omega) v\| = \gamma_1(\alpha)  , \nonumber\\
& &\forall v \in {\cal V} \setminus \{0\}  \enspace,
\lim_{n \rightarrow +\infty} \frac{1}{n} \log \|\tau(T(1,{S^{n} \omega}))
\ldots \tau(T(1, \omega)) v\| = -\gamma_1(\alpha) .
\eea
The properties  (\ref{morphisme}), (\ref{413}) and (\ref{continuite}) imply that 
$\forall v \in {\mathbb R}^4$,
\bea
& &\lim_{n\ra\infty}\frac{1}{n} \log \|\tau(T(1,{S^{n} \omega})) \ldots 
\tau(T(1, \omega)) v\|\nonumber\\
& &= \lim_{n\ra\infty}\frac{1}{n} \log \| T(1,{S^{n} \omega}) \ldots 
T(1, \omega) \rho^{-1}(v)\|.
\eea
Let $v_0 \in {\cal V}$, $u_0 =\rho^{-1}(v_0)$ and ${\cal V}_0 =
{\mathbb C} u_0$. The equation above proves the first assertion.
Consider $u \in {\mathbb C}^2
\setminus {\cal V}_0$. Then 
\bea
u&=& \alpha u_0 + u_0^{\perp},\enspace u_0^{\perp} \neq
0, \enspace \alpha \in {\mathbb C}\\
\rho(u)&=& \Re(\alpha)\rho(u_0) + \Im(\alpha)\rho(i u_0) +
\rho(u_0^{\perp}) \enspace .
\eea
The three components are non zero and mutually orthogonal. Therefore
$\rho(u) \in {\mathbb R}^4 \setminus {\cal V}$, from which the second 
assertion follows.

We can proceed along the same lines in order to prove the 
statements concerning the existence and properties of 
a splitting of $\C^2=E_2^0(\omega)\oplus E_1^0(\omega) $
with $E_j^0(\omega)=\C \rho^{-1}(v_j)$, $v_j\in E_j(\omega)$. Indeed,
let $v_1\in E_1(\omega)$ and $u_1=\rho^{-1}(v_1)$. We define 
$v_1'=\rho(iu_1)$, so that $\bra v_1 | v_1'\ket =0$ and 
$\lim_{n\ra\pm\infty} \frac{1}{n}\ln \|\Phi(n,\omega)v_1'\|=\lim_{n\ra\pm\infty} 
\ln \frac{1}{n} \|\Psi(n,\omega)iu_1\|=\gamma_1$. Hence $v_1'\in E_1(\omega)$. Let
$v_2\in E_2$ such that $u_2:=\rho^{-1}(v_2)$ is not collinear to $u_1$. 
There exists such a $v_2$, otherwise, $u_2=\alpha u_1$ implies 
$\rho(u_2)=v_2=\Re(\alpha)v_1+\Im(\alpha)v_1'\in E_2$, which is a contradiction.
Hence, $v_2'=\rho(iu_2)\in E_2$. Now $u=\alpha u_1+\beta_1 u_2$ and
$\rho(u)=\Re(\alpha)v_1+\Im(\alpha)v_1'+\Re(\beta)v_2+\Im(\beta)v_2'$. So
that 
\be
\lim_{n\ra\pm\infty} \frac{1}{n}\ln \|\Phi(n,\omega)u\|=\gamma_j=
\lim_{n\ra\pm\infty} \frac{1}{n}\ln \|\Psi(n,\omega)\rho(u)\|
\ee
is equivalent to $\beta=0$ if $j=1$ and $\alpha=0$ if $j=2$.
\hfill \ep

\subsection*{Positivity of the Lyapunov exponent}

In order to assess the positivity of the first Lyapunov exponent,
we use Furstenberg's Theorem. Let us introduce, according to
 \cite{bl} III.2.1., the following notions. 

 Let  $S$ be a subset of  $GL_d ({\mathbb R})$, $d>0$. 
Such a set $S$ is said  {\em irreducible} if there is no 
strict subspace $V$ of  ${\mathbb R}^d$ such that
 $\forall M \in S$, $M(V) =V$. A set will be called {\em strongly 
  irreducible} if there is no finite family $V_1, \ldots, V_N$ of strict 
subspaces of ${\mathbb R}^d$, such that: 
$\forall M \in S, M(V_1 \cup \ldots \cup V_N) = V_1 \cup \ldots
  \cup V_N$.

The basic theorem is then
\begin{thm}\label{furstenberg}(Furstenberg)
If $\mu$ is a probability measure on ${\cal M}=\{M \in GL_d ({\mathbb
  R}); |\det M| =1\}$ such that:\\
 $\int \log \|M\| d\mu(M) < +\infty$ and the group ${\cal G}_{\mu}$ 
 generated by the support of $\mu$ is strongly irreducible and non-compact.

Then the first Lyapunov exponent associated with any sequence of i.i.d.
matrices in ${\cal M}$ satisfies $\gamma_1 > 0$.
\end{thm}
See \cite{bl} theorem III.6.1 for a proof.

We note the following property (exercise IV.2.9 of \cite{bl}) 
reducing strong irreducibility to irreducibility in some cases. 
\begin{lem}\label{exo} Let $1<d \in {\mathbb N}$ and 
  ${\cal S}$ a connected subset of $GL_d ({\mathbb R})$. 
 Then ${\cal S}$ is strongly 
  irreducible if and only if ${\cal S}$ is irreducible.
\end{lem}

In our case, the measure $\mu$ is the image by the map (\ref{mapt})
of the uniform measure $P$ on $\T^2$. 
In order to study the properties of
the corresponding set ${\cal G}_{\mu}$, we introduce the connected 
set of matrices given by
the range of the smooth map from $\T^2\rightarrow \C^2$ which to
$(\theta, \eta)$ assigns the matrix
\be\label{redmap}
 T_{(\theta, \eta)} = \left( \begin{array}{cc}
    -e^{-i \theta} & \frac{ir}{t}
    (e^{-i \theta} -1)\\
-\frac{ir}{t} (e^{-i \theta}
-e^{i(\eta -\theta )}) & -\frac{r^2}{t^2}(
e^{-i \theta} -1 - e^{i (\eta -\theta)})-\frac{1}{t^2}
e^{i \eta} \end{array} \right).
\ee
Let ${\cal G}\subset {\cal G}_{\mu}$ denote the smallest group generated
by the support of the measure image by $P$ on 
${SL}_4({\mathbb R})$ by $(\theta, \eta) \rightarrow \tau(T_{(\theta, \eta)})$. 

\begin{prop} The group ${\cal G}$ is not compact.
\end{prop}
{\bf Proof: } The matrix $\tau (T_{(\pi, \pi)})$ belongs to the support 
of the image of  $P$ by (\ref{redmap}) and it has
eigenvalues
\be
\frac{(r-1)^2}{t^2} \mbox{  and  } \frac{(r+1)^2}{t^2}.
\ee
The second one is strictly larger than $1$, if $t<1$ so that, 
since for any $n\in \N$, $\tau (T_{(\pi, \pi)})^n \in {\cal G}$, 
${\cal G}$ cannot be bounded. \ep

\begin{prop} The group ${\cal G}$ is strongly irreducible.
\end{prop}
{\bf Proof: } It is enough to exhibit an irreducible, connected, 
subset of ${\cal G}$. The map $\tau(T_{\cdot, \cdot})$ is smooth, hence
the set $\{\tau(T_{(\theta, \eta)}), (\theta, \eta) \in [0, 2 \pi [^2\}$, 
included in ${\cal G}$, is connected. 
We now show that there exists no strict subspace of $\R^4$ invariant
under this set of matrices.
We first note that choosing $\eta=\theta \in [0, 2\pi [$ we get
\be
 \tau(T_{(\theta, \theta)}) = 
 M_0 + \sin (\theta) M_1+\cos (\theta) M_2
\ee
where
\bea
M_0 &=& \left( \begin{array}{cccc} 0 & 0 & 0 & -\frac{r}{t}\\
    0 & 0 & \frac{r}{t} & 0\\ 0 & \frac{r}{t} & 2\frac{r^2}{t^2} & 0 \\
    -\frac{r}{t} & 0 & 0 & 2\frac{r^2}{t^2} \end{array} \right) \enspace,\\
M_1 &=& \left( \begin{array}{cccc} 0 & 1 & \frac{r}{t} & 0\\
    -1 & 0 & 0 & \frac{r}{t}\\ -\frac{r}{t} & 0 & 0 & -1\\
    0 & -\frac{r}{t} & 1 & 0 \end{array} \right)\enspace, \\
M_2 &=& -(M_0+\un\,),
\eea 
where $\un$ denotes the identity matrix.
If there exists a strict invariant subspace ${\cal E}$ for the 
set  $\tau(T_{\theta})_{\theta \in [0, 2\pi [}$, this subspace
${\cal E}$ is also invariant for the matrices $M_j$, $j=0,1,2$.
Similarly, choosing $-\eta=\theta \in [0, 2\pi [$, we have
\be
 \tau(T_{(\theta, -\theta})) = 
 N_0 + \sin \theta N_1+\cos \theta N_2+\sin (2\theta) N_3+\cos (2\theta) N_4,
\ee
where, in particular,
\be
N_1=\left( \begin{array}{cccc} 0 & 1 & \frac{r}{t} & 0\\
    -1 & 0 & 0 & \frac{r}{t}\\ -\frac{r}{t} & 0 & 0 & \frac{r^2+1}{t^2}\\
    0 & -\frac{r}{t} & -\frac{r^2+1}{t^2} & 0 \end{array} \right)\enspace .
\ee
Again ${\cal E}$ must be invariant under $N_1$.

As $M_0, M_1, N_1$ are real (anti) symmetric, they all leave
${\cal E}^{\perp}$  invariant as well so that these matrices are reduced
by the orthogonal spaces ${\cal E}\oplus {\cal E}^{\perp}=\R^4$. In particular,
these invariant subspaces must be generated by the eigenvectors 
$\{u_1, u_2, u_3, u_4\}$ of $M_0$ which form a basis of $\R^4$. Explicitly,
\be
u_1 = \left( \begin{array}{c} 1\\ 0 \\ 0 \\-\frac{r+1}{t} 
  \end{array} \right),
u_2 = \left( \begin{array}{c} 0 \\ \frac{1-r}{t} \\1 \\0
  \end{array} \right),
u_3 = \left( \begin{array}{c} 1 \\0  \\0 \\ \frac{1-r}{t}
  \end{array} \right) ,
u_4 = \left( \begin{array}{c} 0 \\ -\frac{r+1}{t} \\1 \\0
  \end{array} \right) ,
\ee
the first two vectors being associated with the eigenvalue 
$r(r+1)/t^2$ while the last two are associated with $r(r-1)/t^2$.
We further compute, repeatedly using $r^2+t^2=1$, that
\be
M_1u_1=\frac{1}{t}u_4,\,\,  M_1u_2=\frac{1}{t}u_3,\,\,  M_1u_3=-\frac{1}{t}u_2,\,\, 
M_1u_4=-\frac{1}{t}u_1 
\ee
and
\be
N_1u_1=-\frac{1+r}{t(1-r)}u_2,\,\,  N_1u_2=\frac{1}{t}u_1,\,\,  
N_1u_3=\frac{1-r}{t(1+r)}u_2,\,\, N_1u_4=-\frac{1}{t}u_3. 
\ee
Clearly no one dimensional subspace ${\cal E}=<u_j>$ 
(or  ${\cal E}^{\perp}=<u_j>$) can
be invariant under $M_0, M_1$ and $N_1$. And by inspection, 
one checks that no two dimensional subspace ${\cal E}=<u_j, u_k>$ can be
invariant under $M_0, M_1$ and $N_1$. The irreducible set 
$\{\tau(T_{(\theta, \eta)}), (\theta, \eta) \in [0, 2 \pi [^2\}$ being contained
in the group ${\cal G}$, the latter and  ${\cal G_{\mu}}$ are 
{\em a fortiori} irreducible. \ep

Therefore, 
\begin{prop}\label{poslya}
The Lyapunov exponent $\gamma_1(\lambda)$ associated to the ergodic
linear dynamical system (\ref{rlds}) is strictly positive for any
$\lambda\in\T$.  
\end{prop}

\subsection*{Ishii-Pastur}
The link between Lyapunov exponents and a.c. spectrum is provided 
in the self adjoint random case by the Ishii-Pastur-Kotani Theorem.
We provide a unitary version of the Ishii-Pastur part of the result,
which is enough for our purpose. In order to adapt the proof of 
\cite{cfks}, it is only necessary to show that it is spectrally
true that the generalized eigenvectors of $U$ are polynomially bounded.

We first show that generalized eigenvectors corresponding to
spectral parameters outside the spectrum cannot be polynomially 
bounded for bounded normal operators with a band structure.
We'll say that a matrix $\{M_{j,k}\}_{j,k\in \Z}$ has a 
{\em band structure of order $2p+1$, $p\in\N$} if $|j-k|>p$ implies 
$M_{j,k}=0$. Note that if this is so, then
\be
(M v)_k=\sum_{j\in\Z}M_{k,j}v_j=\sum_{j=k-p}^{k+p}M_{k,j}v_j
\ee
makes sense for an {\em arbitrary} vector $v=\{v_j\}_{j\in\Z}$, since 
the sum is  finite. Define the projections 
\be
  P_{[a,b]}=\sum_{a\leq j\leq b}|\ffi_j\ket\bra \ffi_j |
\ee
and note that
\bea\label{445}
 P_{[a,b]}U&=& P_{[a,b]}U P_{[a-p,b+p]}\nonumber\\
UP_{[a,b]}&=& P_{[a-p,b+p]}U P_{[a,b]}.
\eea
That is, in fact, just another way of saying that $U$ has band structure.
\begin{lem}\label{ptresolv}
Let $(\ffi_n)_{n \in {\mathbb Z}}$ be an orthonormal basis of
a separable Hilbert space ${\cal H}$ on which a normal operator
$U$ acts. Assume $U$ has a band structure of order $2p+1$ and
consider an arbitrary nontrivial sequence $\phi$ such that $U \phi = z \phi$
where $z\in \C$ is in the resolvent set of $U$. Then the sequence
$(\bra \ffi_k| \phi \ket)_{k \in {\mathbb Z}}$ is not polynomially bounded.
\end{lem}
{\bf Proof: } The operator $U$ being normal, for any $z$ in the resolvent set, 
$(U-z)^{-1}$ is normal too.  Therefore 
$\| (U-z)^{-1} \| = r_{\sigma} ((U-z)^{-1})$,
where $r_{\sigma}(A)$ is the spectral radius of the operator $A$.  
As
$r_{\sigma} ((U-z)^{-1})=1/\mbox{dist }(z,\sigma(U))$ (\cite{k} 
(III  6.16 p.177)), we deduce
\be\label{c4}
\forall \psi \in {\cal H}, \enspace \|\psi\| \leq \frac{1}{\mbox{dist }(z,
  \sigma(U))} \|(U-z) \psi \| \enspace .
\ee
Consider the generalized eigenvector $\phi$. Since $z\notin \sigma(U)$,
$\phi$ cannot be in $l^2$, so it must fail to be in $l^2$ either at
$+\infty$ or at $-\infty$. We will assume that it fails at $+\infty$, and
focus on the coefficients $\bra \ffi_k| \phi \ket$, with $k\geq 0$
and large enough. 

Let $n>3p$ and let 
\be
  P_n=P_{[p,n]}=\un -Q_n.
\ee
Since $P_n\phi \in l^2(\Z)$, we have by (\ref{c4})
\be\label{jm}
\|P_n \phi\| \leq c_z \|(U-z)
P_n\phi \| 
\ee
where $c_z^{-1}=\mbox{dist }(z, \sigma(U))$. Since we have assumed that
$\phi$ is not in $l^2$ as $k\geq 0$, necessarily 
\be
  \|P_{n-p}\phi\|^2=\|P_{[p, n-p]}\phi\|^2  \ra \infty \mbox{ as }n\ra \infty.
\ee
So there exists an $n_0$ such that, given $\epsilon >0$,
\be\label{447}
\|P_{n-p}\phi\|^2 \geq \epsilon^{-1}\|P_{[0, 2p]}\phi\|^2.
\ee
Since $(U-z)\phi=0$, it follows for any finite projection $P$ that
\be
(U-z)P\phi=-(U-z)Q\phi
\ee
where $Q=\un -P$, and hence that
\bea
  (U-z)P\phi&=&P(U-z)P\phi+Q(U-z)P\phi\nonumber\\
&=&-P(U-z)Q\phi+Q(U-z)P\phi\nonumber\\
&=& -PUQ\phi +QUP\phi.
\eea
Take in (\ref{jm}), $P=P_n=P_{[p, n]}=\un -Q_n$. By (\ref{445}), we get
\bea
P_nUQ_n&=& P_{[p, n]}UP_{[0, p+n]}Q_n\nonumber\\
&=& P_{[p, n]}U(P_{[0, p-1]}+P_{[n+1, p+n]}).
\eea
Also,
\be
Q_nUP_n = Q_nU(P_{[p, 2p]}+P_{[2p+1, n-p]} +P_{[n-p+1, n]}).
\ee
But 
\be
  Q_nU P_{[2p+1, n-p]}= Q_n P_{[p+1, n]}U P_{[2p+1, n-p]}=0
\ee
so that 
\be
 Q_nUP_n =Q_nU(P_{[p, 2p]} +P_{[n-p+1, n]}).
\ee
Since the ranges of the appropriate projectors are orthogonal,
we have with $A=\|U\|^2$,
\bea
  &&\|(U-z)P_n\phi\|^2=\|P_nUQ_n\phi\|^2 + \|Q_nUP_n\phi\|^2\nonumber\\
&&=\|P_nU(P_{[0,p-1]} +P_{[n+1,p+n]})\phi\|^2
+\|Q_nU(P_{[p,2p]}+P_{[n-p+1,n]})\phi\|^2\nonumber\\
&&\leq A\left( \| P_{[0,p-1]} \phi\|^2+\| P_{[n+1,p+n]} \phi\|^2 
+ \| P_{[p,2p]} \phi\|^2+\| P_{[n-p+1,n]} \phi\|^2 \right)\nonumber\\
&&= A\left( \| P_{[0,2p]} \phi\|^2+\| P_{[p,n+p]} \phi\|^2 
-\| P_{[p,n-p]} \phi\|^2 \right)\nonumber\\
&& = A\left( \| P_{[0,2p]} \phi\|^2+\| P_{n+p} \phi\|^2 
-\| P_{n-p} \phi\|^2 \right).
\eea
Thus, by (\ref{447}) and (\ref{jm}), for $n>\max (n_0,3p)$, we have
\bea
  \|P_{n-p}\phi\|^2&\leq&\|P_n\phi\|^2\leq c_z^2 \|(U-z)\phi\|^2\nonumber\\
&\leq&c_z^2 A\left(\epsilon \|P_{n-p}\phi\|^2+\|P_{n+p}\phi\|^2
-\|P_{n-p}\phi\|^2 \right),
\eea
which implies that
\be
  \|P_{n+p} \phi\|^2 \geq \|P_{n-p} \phi\|^2(\frac{1}{Ac_z^2}+1-\eps))\equiv 
B\|P_{n-p} \phi\|^2,
\ee
where $B>1$, if $\epsilon< 1/(Ac_z^2)$.
Iterating the argument, we get
 $\forall k \in {\mathbb N},$
\be
\| P_{n+p2k} \phi \| \geq B^{\frac{k}{2}} \| P_n
\phi\|\enspace ,
\ee
which ensures the existence of an exponentially growing subsequence
of coefficients. 
\hfill \ep\\
The second element is the construction of generalized solutions 
corresponding to spectral parameters in the spectrum of $U$ which
are polynomially bounded, {\sl \`a la} Berezanskii. 
This is done in our unitary setting following,  
{\sl mutatis mutandis}, the arguments 
given in \cite{s} for the self-adjoint case. We only quote the end 
result here, including a proof in Appendix for completeness.

Recall that a measure $\rho$ is in the measure class of a unitary 
operator $U$ with spectral projection $E(\cdot )$ if 
$\rho(\Delta)=0 \Leftrightarrow E(\Delta)=0$ for any Borel set $\Delta$.
\begin{thm}\label{solpoly}
Let $U$ be a unitary operator with a band structure defined on 
$l^2({\mathbb Z})$ and 
$\delta > 1$. Then there exists a measure $\rho$ in the spectral measure 
class of $U$ and a family of disjoint measurable sets 
$(\Delta_n)_{n \in {\mathbb N}^*}$ whose union supports $\rho$ such that
for $\lambda \in \Delta_n$, there exist $n$ vectors  
$\phi_{j}(\lambda)$  satisfying
\begin{itemize}
\item $(U-e^{i\lambda})\phi_j(\lambda)=0$ 
\item $\forall n \in {\mathbb Z}, |\bra\ffi_n|\phi_j(\lambda)\ket| \leq C <n>^{\delta}$
\item For any $\lambda$ fixed, the family $\{\phi_j(\lambda)\}_j$ is 
linearly independent.
\end{itemize}
\end{thm}
{\bf Remark:} The result is also true if the operator $U$ is defined
on $l^2(\N)$.
\begin{cor}\label{simonrevu} $\sigma(U)$ is the closure of the set 
\be
S = \{ \lambda \in {\mathbb T}^1; U\phi =e^{i\lambda}\phi \mbox { admits 
a polynomially bounded solution} \}
\ee 
and 
$E_{[0,2\pi[ \setminus S} (U) =0$.
\end{cor}
{\bf Proof: }  
If $e^{i\lambda} \in \sigma(U)$ then for any $\epsilon
>0$, $E(]\lambda -\epsilon, \lambda +\epsilon[) > 0$ and
$\rho(]\lambda -\epsilon, \lambda +\epsilon[) > 0$. Hence, 
by Theorem \ref{solpoly}, for $\lambda'$ arbitrarily close to 
$\lambda$ there exists a polynomially bounded 
solution $\phi_j(\lambda')$. 
Thus $ \sigma(U) \subset \bar{S}$. The reverse inclusion follows from
Lemma \ref{ptresolv} and the fact that $\sigma(U)$ is closed.
The last statement follows immediately.
\ep \newline

Putting these arguments together, we get 
the unitary version of the Ishii-Pastur theorem suited to our 
monodromy operator:
\begin{thm}\label{IP}
Let $U_{\omega}$ be unitary with a band structure. Assume
that the corresponding transfer matrix at spectral parameter 
$\E^{i\lambda}$ induces two Lyapunov exponents
$\gamma_1(\lambda)\geq \gamma_2(\lambda)=-\gamma_1(\lambda)$ 
which are constant almost surely. Then
\be
\sigma_{ac}(U_{\omega}) \subseteq \overline{ \{ e^{i \lambda} 
\in S^1 ; \gamma_1 (\lambda)=0 \} } \enspace .
\ee
\end{thm}
{\bf Proof:} Identical to that given in \cite{cfks} Thm 9.13.
\ep

Therefore, Theorem \ref{main} follows from the above Theorem
and Proposition \ref{poslya}

\setcounter{equation}{0}
\section{Coherent setting}\label{coher}

In that section we consider situations where the behavior of the matrix 
coefficients of $U$ in (\ref{matel}) are periodic functions of $k$ as 
the result of a coherent behavior of the phases. We first show that this
implies purely absolutely continuous spectrum. Then we prove that when restricted 
to $l^2(\N)$, these operators have no singular continuous spectrum
and may possess finitely many simple eigenvalues only.

\subsection*{Coherence on $l^2(\Z)$}

As a first particular case of coherent dependence of the scattering phases, 
we consider the simple situation where the $\theta_k$'s and $\alpha_k$'s 
take alternatively two values, up to a linear term. This corresponds 
to a monodromy operator $U=U_oU_e$ where $U_e$, $U_o$ are direct sums of 
constant blocks $S_{2k}=S_e, \,\,S_{2k+1}=S_o$.
\begin{prop} 
Let $t\in]0,1[$, let the sequence $\{\gamma_k\}$ be arbitrary and
$$
\theta_k=\left\{\matrix{\theta_e \mbox{ if }k\mbox{ is even}\cr
\theta_o \mbox{ if }k\mbox{ is odd }}\right.\,\,\, ,
\alpha_k=ak+\left\{\matrix{\alpha_e \mbox{ if }k\mbox{ is even}\cr
\alpha_o \mbox{ if }k\mbox{ is odd }}\right.\,\, \forall k\in\Z,
$$
where $\theta_e, \theta_o, \alpha_e, \alpha_o, a\in\R$. Define
$\Delta=\alpha_e-\alpha_o$, $\Theta=\theta_e+\theta_o$.
Then, with the identification $U\equiv M(\{\theta_k\},\{\alpha_k\},\{\gamma_k\})$, 
$U$ is purely absolutely continuous and  
$$\sigma_{ac}(U)=\{\E^{-i(a+\Theta)}
\E^{\pm i(\arccos(r^2\cos\Delta -t^2\cos(2x+\Delta)))}, x\in\T\} $$
\end{prop}
{\bf Proof:} By Lemma \ref{free}, we can replace $\gamma_k$ by
$(-1)^{k+1}\alpha_k$ so that, with our choice of phases, 
$U\simeq \E^{-i(a+\Theta)}V$, where 
\bea\label{matelv}
V\ffi_{2k}&=&irt\E^{-i\Delta}\ffi_{2k-1}+r^2\E^{-i\Delta}\ffi_{2k}
+irt\E^{i\Delta}\ffi_{2k+1}-t^2\E^{i\Delta}\ffi_{2k+2}\nonumber\\
V\ffi_{2k+1}&=&-t^2\E^{-i\Delta}\ffi_{2k-1}
+itr\E^{-i\Delta}\ffi_{2k}
+r^2\E^{i\Delta}\ffi_{2k+1}
+irt\E^{i\Delta}\ffi_{2k+2}.\nonumber\\
& &
\eea
Let us map $l^2(\Z)$ unitarily to $L^2(\T)$ via
\be 
W: \ffi_k \mapsto \E^{ikx},
\ee
such that for any $\psi=\sum_k c_k\ffi_k$, $c_k\in\C$ 
\be
 (W\psi)(x)=\sum_{k\in\Z} c_k \E^{ikx}\in L^2(\T).
\ee
We further introduce $L^2_{\pm}(\T)=WL_{\pm}$, where $L_{+},L_{-} $ are 
the subspaces of $l^2(\Z)$ generated by the basis vectors with even, 
respectively odd, indices. 
It is easily checked that
$V$ is unitarily equivalent on $L^2(\T)=L^2_{+}(\T)\oplus L^2_{-}(\T)$ 
to the matrix valued multiplication operator 
\be
V\simeq \pmatrix{r^2\E^{-i\Delta}-t^2\E^{i\Delta}\E^{2ix} & 2itr\cos(x+\Delta)\cr
2itr\cos(x+\Delta)& r^2\E^{i\Delta}-t^2\E^{-i\Delta}\E^{-2ix}}.
\ee
This matrix is analytic in $x$ and has non-constant eigenvalues given 
by 
\be
\lambda_{\pm}(x)=r^2\cos\Delta-t^2\cos(\Delta+2x)\pm i\sqrt{1-(r^2\cos\Delta-t^2\cos(\Delta+2x))^2},
\ee
from which the result follows.\hfill \ep

Using basically the same strategy, we can consider the general case 
where the elements of $U$ display an arbitrary periodicity.
\begin{thm}\label{thco}
Let $t\in]0,1[$, let the sequence $\{\gamma_k\}$ be arbitrary and
$\{\theta_k\},\{\alpha_k\}$ be such that for some $2\leq N\in\N$, 
and all $k\in\Z$,
$$\theta_{k+N}=\theta_k, \,\,\,\, \alpha_{k}=ak+\pi_k, \,\,\mbox{ where }\,\,
\pi_{k+N}=\pi_k \,\,\mbox{ and }\,\, a\in\R.$$
Then, with the identification $U\equiv M(\{\theta_k\},\{\alpha_k\},\{\gamma_k\})$, $U$ is purely absolutely
continuous.
\end{thm}
{\bf Proof:}  As above, we first replace $\gamma_k$ by $(-1)^{k+1}\alpha_k$ 
and we introduce 
\bea\label{decl2}
L^2({\mathbb T}) &= \left[ \oplus_{q=0}^{N-1} L_{2q}^2({\mathbb T}) \right] \bigoplus
\left[ \oplus_{q=0}^{N-1} L_{2q+1}^2({\mathbb T}) \right] \\
L_{2q}^2({\mathbb T}) &= \overline{\mbox{span} \{(e^{(2Nk + 2q)i x })_{k \in {\mathbb
      Z}}, x\in\T \}}\\
L_{2q+1}^2({\mathbb T}) &= \overline{\mbox{span} \{(e^{(2Nk +2q +1)i x})_{k \in {\mathbb
      Z}}, x\in\T \}} \enspace .
\eea
If $P_{2q}$ and $P_{2q+1}$ denote the orthogonal projections on these 
subspaces, we get for $\psi=\sum_{k\in\Z}c_k\ffi_k$, with the same 
notations as above, 
\be
 (P_{2q}W\psi)(x)=\sum_{k\in\Z}c_{2(Nk+q)}\E^{i2(Nk+q)x}\in L^2_{2q}(\T),
\ee
and similarly for $(P_{2q+1}W\Psi)(x)$. To determine the image of $U$ under
the unitary mapping $W$, we introduce
\be
\nu_{k}^{\pm}=\theta_{2k}+\theta_{2k\pm 1}\mp(\pi_{2k}-\pi_{2k\pm 1}),
\ee
such that $\nu_{k}^{\pm}=\nu_{k+N}^{\pm}$ for any $k\in\Z$. Hence
$U\simeq \E^{-ia}V$ where, this time, $V$ acts on $l^2(\Z)$ according to 
\bea\label{matelper}
V\ffi_{2k}&=&irt\E^{-i\nu_{k}^-}\ffi_{2k-1}+r^2\E^{-i\nu_{k}^-}\ffi_{2k}
+irt\E^{-i\nu_{k}^+}\ffi_{2k+1}-t^2\E^{-i\nu_{k}^+}\ffi_{2k+2}\nonumber\\
V\ffi_{2k+1}&=&-t^2\E^{-i\nu_{k}^-}\ffi_{2k-1}
+itr\E^{-i\nu_{k}^-}\ffi_{2k}
+r^2\E^{-i\nu_{k}^+}\ffi_{2k+1}
+irt\E^{-i\nu_{k}^+}\ffi_{2k+2}.
\eea
The phases $\nu_{k}^{\pm}$ being $N$-periodic, by manipulations 
similar to those performed above, one gets that 
$V\simeq T$ where $T$ is a matrix valued multiplication operator on
the decomposition of the Hilbert space (\ref{decl2})
by the $2N\times 2N$ matrix
\be\label{matval}
T(\E^{ix})=\sum_{k=-2}^2\E^{ikx}T_k
\ee
where the $T_k$'s have a $N\times N$ block structure of the form
\bea\label{613}
&&T_{2}=-t^2\pmatrix{0 & 0 \cr 0 & W_u}, T_{1}=
irt\pmatrix{0 & D_- \cr W_u& 0}, \\
&&T_0=r^2\pmatrix{D_- &0 \cr 0 & D_+}, T_{-1}=irt\pmatrix{0 & W_l \cr D_+& 0}, 
T_{-2}=-t^2\pmatrix{W_l & 0 \cr 0 & 0},\nonumber
\eea
with
\bea
&&D_{\pm}=\mbox{diag}(\E^{-i\nu_0^{\pm}}, \E^{-i\nu_{1}^{\pm}}, \cdots , 
\E^{-i\nu_{N-1}^{\pm}})\nonumber\\
&&W_u=\pmatrix{0 & \E^{-i\nu^-_{1}}& & & \cr &0 &\E^{-i\nu^-_{2}} & & \cr & &\ddots &  & 
\cr & & &0 & \E^{-i\nu^-_{N-1}}
\cr
\E^{-i\nu^-_{0}} & & & &0},
\nonumber\\
&&W_l=\pmatrix{0 & & & &\E^{-i\nu^+_{N-1}} \cr\E^{-i\nu^+_{0}} &0 & & & \cr
 & \E^{-i\nu^+_{1}}&\ddots &  & \cr &  & &0&  
\cr & & & \E^{-i\nu^+_{N-2}}&0}.
\eea
Now, the operator $T$ being unitary, the matrix $T(\E^{ix})$ is 
unitary as well for almost every $x\in\R$. But this matrix 
being analytic in a neighborhood of the real axis, it must be
unitary everywhere on the real axis. By classical results 
in analytic perturbation theory, see \cite{k}, it is therefore 
diagonalizable with analytic eigenprojectors in a neighborhood 
of the real axis, and identically zero eigennilpotents.
In order to prove the absolutely continuous nature of the spectrum of $U$,
it is then enough to show that the analytic eigenvalues of the matrix 
$T(\E^{ix})$ are non constant in $x\in \R$. But this is immediate, because
otherwise, an infinitely degenerate eigenvalue would exist, which is forbidden
by (\ref{cocycle}).\ep\\ \\
{\bf Remarks:} \\
i) The formulae (\ref{613}) above are the starting point of a detailed
analysis of the band spectrum of $U$ as a function of $t\in ]0,1[$, which we shall
not perform here. We only note that for $t=0$
\be
  \sigma(U)=\{\E^{-i\nu_0^{+}}, \E^{-i\nu_{1}^{+}}, \cdots , 
\E^{-i\nu_{N-1}^{+}}, \E^{-i\nu_0^{-}}, \E^{-i\nu_{1}^{-}}, \cdots , 
\E^{-i\nu_{N-1}^{-}}\},
\ee
where each eigenvalue is infinitely degenerate, whereas, for
$t=1$,
\bea
  \sigma(U)=\bigcup_{k=0,\cdots, N-1}&&\mbox{Ran}\{\E^{-2ix}\E^{-i(\nu_0^{+}+
\nu_{1}^{+}+\cdots \nu_{N-1}^{+})/N}
\E^{ik2\pi/N}, x\in\T\}\nonumber\\
&&\cup \,\,\,\,\mbox{Ran} \{\E^{2ix}\E^{-i(\nu_0^{-}+\nu_{1}^{-}+
\cdots \nu_{N-1}^{-})/N}\E^{ik2\pi/N}, x\in\T\}.
\eea
Perturbation theory as $t\ra 0$ and $t\ra 1$ can now be applied to get 
information on the corresponding band functions in these regimes.\\
ii) It is not difficult to check that a unitary band matrix of order 
$2p+1$ with periodic coefficients, in the sense that there exists $N>0$ such that 
$\bra\ffi_j |U\ffi_k\ket =\bra\ffi_{j+N} |U\ffi_{k+N}\ket$, is always unitarily
equivalent to a multiplication by an $pN\times pN$ unitary matrix $T(\E^{ix})$ on 
$\oplus_{q=0,\cdots, pN-1}L_q^2(\T)$, where $T(\E^{ix})$ is a polynomial of degree
$p$ in $\E^{\pm ix}$. However, in general, one cannot rule out the existence 
of finitely many infinitely degenerate eigenvalues.

\subsection*{Coherence on $l^2(\N^*)$}

Let us now turn to the study of $U^+$ defined on $l^2(\N^*)$ by (\ref{defup})
in case the phases $\{\gamma_k\}$ are arbitrary whereas $\{\theta_k\}$ and $\{\alpha_k\}$ are eventually 
coherent: {\it i.e.}, there exists $k_0\in\N$ and
$2\leq N\in\N$ such that for all $k\geq k_0\in \N^*$,
\be\label{cohbe}
\theta_{k+N}=\theta_k, \,\,\,\, \alpha_{k}=ak+\pi_k, \,\,\mbox{ where }\,\,
\pi_{k+N}=\pi_k \,\,\mbox{ and }\,\, a\in\R.
\ee
We can replace without loss $\gamma_k$ by $(-1)^{k+1}\alpha_k$ and assume
$a=0$, since we are working up to unitary equivalence.
Our coherent comparison operator $U_0$ on $l^2(\Z)$ is defined by (\ref{matel}) with phases $\{\theta_k\}$ and $\{\alpha_k\}$ obtained by extending  
(\ref{cohbe}) (with $a=0$) to $\Z$. Therefore
we can write on $l^2(-\N^*)\oplus \C \oplus l^2(\N^*)$
\be
U_0= \pmatrix{W^- & &
\cr &1& \cr
& & U^+} -F
\ee 
where absent elements denote zeros, $W^-$ is an operator
defined on $l^2(-\N^*)$  which is eventually periodic 
and $F$ is a finite rank operator. It is always possible to 
construct $U_0$ this way with $\dim \mbox{ Ran }F=M$ depending on
$N$ and $k_0$.
\begin{thm} Let $U^+$ and $U_0$ be as above.
Then $$\sigma_{s.c.}(U^+)=\emptyset\ \,\mbox{ and }\, 
\sigma_{a.c.}(U^+)=\sigma_{a.c.}(U_0).$$ 
The point spectrum of $U^+$ consists of finitely many simple eigenvalues 
in the resolvent set of $U_0$.
\end{thm}
{\bf Remark:} As the proof below shows, the same statement holds if $U^+$ 
denotes a doubly infinite coherent matrix perturbed by a finite rank operator.\\  
{\bf Proof:}
Let us first show that 
the finite rank perturbation $F$ of the unitary $U_0$ doesn't produce
any singular continuous spectrum. By Weyl's Theorem, this cannot happen 
in the gaps (on $S^1$) of the absolutely continuous spectrum of $U_0$. 
Therefore we focus on $\sigma(U_0)$.
Depending on $k_0$ and $N$, we have
for some finite $M>0$,
\be\label{finran}
F=\sum_{|j|,|k|\leq M}c_{j,k}|\ffi_j\ket \bra \ffi_k |.
\ee
We know from (\ref{matval}) that $U_0$ is unitarily equivalent
to the multiplication by a $2N\times 2N$ unitary matrix $V(x)$ on 
the decomposition (\ref{decl2}), where $V(x)$ is a polynomial in 
$e^{\pm ix}$ whose eigenvalues are not constant in $x$.  Therefore,
$V(x)$ is analytic in a neighborhood of the real axis and we can 
write 
\be
  V(x)=\sum_{j=1}^{2N}P_j(x)\lambda_j(x)
\ee
where the eigenprojections $P_j$ and eigenvalues $\lambda_j$ are
analytic in a neighborhood of the real axis as well (see \cite{k}).
We know that 
\be
\sigma(U_0)=\cup_{j=1}^{2N}\mbox{ Ran }\{\lambda_j(x)
, \,\, x\in\T\}.
\ee 
Note that 
\be
  F\simeq \sum_{|j|,|k|\leq M}c_{j,k}|(e^{ijx})\ket\bra (e^{ikx}) |
\ee
where the r.h.s is to be understood as a multiplication operator on the
decomposition (\ref{decl2}) and $(e^{ijx})$ is a vector in $\C^{2N}$ with
zero elements except at some line, depending on $j$, where the entry is
$ e^{ijx}$. 
We follow the perturbation theory of unitary operators presented in
\cite{kk} to study the
unitary operator $U_1\equiv U_0+F$.
Let $\zeta=\rho e^{i\beta}$ with $\rho\neq 1$ and $\beta\in \T$. 
We set for $j=0,1$
\bea
R_j(\zeta)&=&U_j(U_j-\zeta)^{-1}=(\un - \zeta{U_j}^*)^{-1}\\
G(\zeta)&=&\un +\zeta(U_1^*-U_0^*)R_1(\zeta)\nonumber\\
&=&(\un +\zeta(U_0^*-U_1^*)R_0(\zeta))^{-1}.
\eea
These quantities are analytic in $\C\setminus S^1$.
We know from \cite{kk} that for any vectors $f, g$,
\be\label{specmes}
 \lim_{\rho\ra 1^-}\bra g |\delta_{\rho}(E_j, \beta) f\ket=
\frac{d}{d\beta}\bra g| E_{a.c., j} (\beta) f\ket \,\,\, 
\mbox{ a.e. }\beta\in\T, \,\,\,j=0,1,
\ee
where 
\be
 2\pi\delta_{\rho}(E_j, \beta)=R_j(\zeta)-R_j(\zeta')\,\, \mbox{ with }
\zeta'=1/\bar{\zeta},
\ee
and $E_{a.c., j} (\beta)$ is the absolutely continuous part of the
the spectral projector of $U_j$ at $e^{i\beta}$. Also,
\bea
\delta_{\rho}(E_1, \beta)&=& G(\zeta)^{*}\delta_{\rho}(E_0, \beta)G(\zeta)
\nonumber\\ &=& {(\un -\zeta F^*R_0(\zeta))^{-1}}^* \delta_{\rho}(E_0, \beta)
 (\un -\zeta F^*R_0(\zeta))^{-1}.
\eea
In order to get information on
the nature of spectral measure of $U_1$, it is sufficient to consider (\ref{specmes}) 
on the cyclic subspace for $U_0$ generated by the range of $F^*$. Indeed, the spectral 
measures of $U_0$ and $U_1$ associated with vectors in the orthogonal complement 
of this subspace coincide and it is cyclic also for $U_1$. Let $P$ denote the projector 
on $\mbox{Ran }F^*$. We first note that $\un -\zeta F^*R_0(\zeta)$ is 
invertible on $\mbox{Ran }P$ if and only if the determinant
\be\label{det}
  \det (P(\un - \zeta F^*R_0(\zeta))P)\neq 0
\ee
and
\be
  (\un -\zeta F^*R_0(\zeta)|_{\mbox{Ran }P})^{-1}=(P(\un -\zeta F^*R_0(\zeta))P)^{-1}P.
\ee
So we need to consider the finite matrix whose elements are given for
$|n|,|m| \leq M$ by
\bea\label{eee}
&&  \bra \ffi_n | F^*R_0(\zeta)\ffi_m\ket =\sum_{|j|\leq M}\bar{c}_{j,n}
\bra\ffi_j |U_0(U_0-\zeta)^{-1}\ffi_m\ket\nonumber\\
&& = \sum_{|j|\leq M}\sum_{l=1}^{2N}\bar{c}_{j,n}\int_{0}^{2\pi}dx
\left<(e^{ijx}), \frac{P_l(x)\lambda_l(x)}{\lambda_l(x)-\zeta}  (e^{imx})\right>,
\eea
where $<\cdot, \cdot >$ denotes here the scalar product in $\C^{2N}$. Therefore,
(\ref{eee}) is a finite sum of the form
\be\label{intex}
  \sum_{l=1}^{2N}\int_{0}^{2\pi}dx \frac{f^{(l)}_{n,m}(x)}{\lambda_l(x)-\zeta},
\ee
where $f^{(l)}_{n,m}$ is analytic in an open strip of finite width, independent
 of $l,n,m$ containing the real axis. 

Fix an $l \in \{1,\cdots, 2N \}$ and let  $x_{\beta}\in\T$ be such that 
$e^{i\beta}=\lambda_l(x_{\beta})$. There is only a finite number of such 
points.
Assume $\lambda_l'(x_{\beta})\neq 0$. Then we can deform the contour of 
integration in $x$ to control the integrals (\ref{intex}) when $\rho\ra 1$ 
as follows. There exists a neighborhood $\C \supset N_{\beta}$ of
$x_{\beta}$ which is mapped by $\lambda_l$ bijectively on its image $M_{\beta}$ which
contains $e^{i\beta}$ in its interior. Let $D_{\beta}\subset M_{\beta}$ be a smooth deformation
of the unit circle towards the exterior which avoids $e^{i\beta}$. Taking the
inverse image $\lambda_l^{-1}(D_{\beta})\subset N_{\beta}$ and connecting it
at both ends with the real axis, we get a smooth path $C_{\beta}$ along which
\be
  \int_{0}^{2\pi}dx \frac{f^{(l)}_{n,m}(x)}{\lambda_l(x)-\zeta}=
\int_{C_{\beta}} dz \frac{f^{(l)}_{n,m}(z)}{\lambda_l(z)-\zeta}.
\ee
By construction, the last integral is now analytic in $\zeta$ in
a neighborhood $\tilde{M_{\beta}}\subset M_{\beta}$ containing  $e^{i\beta}$.
Therefore, the matrix (\ref{eee}) has an analytic continuation in $\zeta$ in a
 neighborhood of $S^1$ except at a finite set of points. Hence there is 
only a finite set of points of $S^1$, call it $Z$, where the determinant 
(\ref{det}) is zero.

Then, for any $\psi =P\psi$ and any $e^{i\beta}\in Z^{C}$, we can write
\be
  (\un -\zeta F^*R_0(\zeta))^{-1}\psi=\sum_{|k|\leq M} d_k(\zeta)\ffi_k,
\ee
where the $d_k(\zeta)$'s are analytic in a neighborhood of $e^{i\beta}$ 
and the $\ffi_k$'s span the range of $F^*$. Thus, we deduce from
the relation
\be
  \bra \psi |\delta_{\rho}(E_1, \beta)\psi \ket=\sum_{|k|, |j| \leq M} \bar{d_j(\zeta)}d_k(\zeta)\bra \ffi_j |\delta_{\rho}(E_0, \beta)
 \ffi_k \ket ,
\ee
that the limit $\rho\ra 1^-$ yields the derivative of an absolutely continuous measure on $S^1\cap Z^{C}$, as the limits $\lim_{\rho\ra 1^-}\bra \ffi_j |\delta_{\rho}({E_0}, \beta) \ffi_k \ket \in L^1(\T)$.  
As a finite set of point cannot support a continuous measure, 
we get that $\sigma_{s.c.}(U_1)=\emptyset$. 

Let us consider the point spectrum of $U^+$. From the relation (\ref{cocycle}),
we get that the eigenvalues have multiplicity two at most.  Except for a finite number of them, the transfer matrices $T(k)$ are periodic in $k$, of period $N$. Therefore we define 
\be
R=T(k_0+1+N)T(k_0+N)\cdots T(k_0+1)
\ee
and set 
\be
d(k)=\pmatrix{c_{2k}\cr c_{2k+1}}
\ee
so that
\be\label{ggg}
d(jN+k_0)=R^jd(k_0)=R^j T(k_0)T(k_0-1)\cdots T(2)d(1).
\ee
We will use the notation  $D(j)=d(jN+k_0)$.
Note also that $\det R=e^{i\kappa}$, where $\kappa\in\T$ is
independent of $\lambda$,
due to (\ref{det=1}), and that the matrix 
$R$ is analytic in $\lambda$ since it is a polynomial in 
$e^{ +i\lambda}$ and $e^{- i\lambda}$.

Assume that an eigenvector of $U^+$ exists in $l^2(\N^*)$ for the
eigenvalue $e^{i\lambda}$. This implies
that the sequence $\{\|D(j)\|\}_{j\in \Z}$ belongs to $l^2$ at $+\infty$.
We are thus lead to the study of (\ref{ggg}). This is done by means of the
following elementary lemma whose proof we omit.
\begin{lem}\label{lemel}
Let $R$ be a $2\times 2$ matrix with $|\det R|=1$, and let $E_1$ be an
eigenvalue of $R$. Consider $D(j)=R^jD(0)$, where $D(0)\in \C^2$. 
Then, \\
1) there exists $K>0$, such that  $\forall D(0)$ of norm 1 and  
$\forall j\in\Z$, $ K \leq \|D(j)\| \leq |j|/K$
if and only if $|E_1|=1$.\\
2) When $|E_1|\neq 1$, there exists another eigenvalue $E_2\neq E_1$. 
We can assume $|E_1|>1>|E_2|=|E_1|^{-1}$ and we get 
$$
D(j)=A E_1^j v_1 +B E_2^j v_2, \,\,\, j\in\Z,
$$
where $v_1, v_2\in\C^2$ are the corresponding eigenvectors of $R$ and
$A,B\in \C$ are the coefficient of $D(0)$ in the basis they form.
\end{lem}
A direct consequence is that $\{\|D(j)\|\}\in l^2(\N)$ implies 
exponential decay at $+\infty$, thus $D(0)=v_2$ and any eigenvalue is simple. 
Now we use $D(0)$ as an initial vector to construct a generalized vector for 
the coherent operator $U_0$ on $l^2(\Z)$.
Note that considered as a functions of $\lambda$, $R(\lambda)$ is analytic in a
 neighborhood of the real
axis, therefore, $E_1(\lambda)$ is analytic on $\T$, except at the finite set 
$X$ of exceptional points in $\T$ where the eigenvalues of $R(\lambda)$ cross.
At such exceptional points, $|E_1|=1$. Then observe that if the second statement of Lemma \ref{lemel} 
is true for some $\lambda\in\T$, it still true in a neighborhood
of $\lambda$ by continuity. This implies that all generalized eigenvectors
corresponding to eigenvalues in the corresponding neighborhood grow
 exponentially at one end or the other.
Due to Corollary \ref{simonrevu}, this
can take place only in the resolvent set of $U_0$. Also, as the spectrum of 
$U_0$ contains no isolated point, the argument above shows that $X$ must belong to the closure of the set of points in $\T\setminus X$ where $|E_1(\lambda)|=1$. Therefore $|E_1|$ is continuous on the whole of $\T$ and 
$\sigma(U_0)=|E_1|^{-1}(\{1\})$. The band edges are also
excluded from the point spectrum of $U^+$, since they correspond 
to points where $|E_1|=1$. 

We now study the number of eigenvalues of $U^+$. The boundary condition
that $d(1)$ has to meet reads, according to (\ref{defup}),
\be
 \tilde{T}^{-1}d(1)=c_1 b(\lambda),
\ee
where $c_1$ is the non zero first coefficient of the generalized eigenvector
and 
\bea
  \tilde{T}^{-1}&=&e^{-i(\theta_1+\theta_2+\alpha_2-\alpha_1)}
\pmatrix{irt & -t^2 \cr r^2 & itr}-e^{i\lambda}\pmatrix{0 & 0 \cr 1 & 0}\\
b(\lambda)&=& e^{i\lambda}\pmatrix{1\cr 0}-e^{-i(\theta_0+\theta_1+\alpha_1)}
\pmatrix{r\cr it},
\eea
with $|\det \tilde{T}^{-1}|=1$. Therefore, the condition to have an eigenvalue
$e^{i\lambda}$ for $U^+$ is equivalent to 
\be
b(\lambda)\,\,\, \parallel \,\,\, \tilde{T}^{-1} T^{-1}(2) T^{-1}(3)\cdots  
T^{-1}(k_0)v_2(\lambda) , 
\ee
where $v_2(\lambda)$ is an eigenvector of $R(\lambda)$ and all matrices involved are analytic in $\lambda\in\T$. In other words, $e^{i\lambda}$
is an eigenvalue if and only if
\be\label{func}
  \det (v_2(\lambda); T(k_0)T(k_0-1)\cdots T(2)\tilde{T}b(\lambda))\equiv
\det (v_2(\lambda);a(\lambda))=0,
\ee
where $a$ is analytic on $\T$ and $v_2$ can be chosen analytic on 
$\T\setminus X$, see \cite{k}. Therefore, to show that the number of eigenvalues
of $U^+$ is finite, it is enough to prove that, as a function of $\lambda$ on $\T$, the determinant above has finitely many zeros. This is a consequence of 
the next Lemma we prove in appendix.
\begin{lem}\label{sqrt}
If $\lambda_0\in X$, the eigenvectors $v_j(\lambda)$, $j=1,2$, have 
 at worst a square root branch point at $\lambda_0$.
\end{lem}
It follows that the function $\lambda\mapsto \det (v_2(\lambda);a(\lambda))$ 
is analytic on $\T\setminus X$
and possesses square root branch points singularities at $X$ as well. Therefore
it only possesses finitely many zeros on $\T$.

Finally, we show that $\sigma(U_0)\subset \sigma(U^+)$. Let
$e^{i\lambda}$ in the interior of the set $\sigma_{a.c}(U_0)$ and consider the relation (\ref{cocycle})
yielding the coefficients $d(k)$ of the corresponding generalized eigenvector.
Up to a finite number of transfer matrices $T(k_0)T(k_0-1)\cdots T(1)$,
this relation is identical to that yielding the coefficients with
positive indices of a corresponding generalized eigenvector for $U^+$.
The discussion above shows that $d(k)$ is polynomially bounded at both
ends, so that by Corollary \ref{simonrevu}, $e^{i\lambda}$ belongs the spectrum
of $U^+$ as well. This finishes the proof of the Theorem. \ep\\ \\
{\bf Remark: } In keeping with the last remark following the proof of
Theorem \ref{thco}, let $U_0$ denote a periodic band matrix of order $2p+1$.
Then, it is also true that a finite rank perturbation of the form (\ref{finran})
produces no singular continuous spectrum, since the first part of the above 
proof goes through without changes.

\setcounter{equation}{0}
\section{An almost periodic example}

In order to complete the picture of the spectral properties
such matrices can possess, we briefly describe below an example 
of deterministic unitary band matrices which is almost periodic and
purely singular continuous. This example is constructed in 
analogy with the random discrete Schr\"odinger case according to
the approaches of Herman and Gordon, see {\sl e.g.} \cite{cfks}.

We consider again the matrix $M(\{\theta_k\}, \{\alpha_k\},\{\gamma_k\})$ 
where the phases $\alpha_k$ are taken as constants, while the $\gamma_k$'s
are arbitrary and can be replaced by $(-1)^{k+1}\alpha_k$, as above.
The almost periodicity lies with the phases $\theta_k$ defined 
according to
\be\label{alpe}
  \theta_k = 2 \pi \beta k+ \theta , \,\,\,\forall k \in {\mathbb N},
\ee
where $\beta$ is irrational, and $\theta\in [0,2\pi[$.

Consider the uniform measure  $P_0$ on the $\T$,
and the translation  $\tau: \T\ra\T$ defined by 
\be
\tau( {\theta}) = 2i \pi  \beta + \theta .
\ee
Then set of iterates $\tau^k$, $k\in\Z$ is ergodic.
The corresponding transfer matrices $T(k)^{\theta}$ generated by this set
of translations are then given by 
(see (\ref{tren}))
\bea\label{tape}
&&T(k)^{\theta}_{11}=  - e^{-i (\lambda+2\theta+8 k \pi \beta-6\beta\pi)} \\
&&T(k)^{\theta}_{12} = \frac{ir}{t} 
(e^{-i (\lambda+2 \theta+8 k \pi \beta-6\beta\pi)} - 1)\nonumber\\
&&T(k)^{\theta}_{21} =\frac{ir}{t}(e^{2i \theta} -e^{-i (\lambda+2 \theta+8 k \pi 
\beta-6\beta\pi)}) \nonumber\\
&& T(k)^{\theta}_{22} =\frac{r^2}{t^2}(e^{4i \beta}+1-
e^{-i (\lambda+2 \theta+8 k \pi \beta-6\beta\pi)})\nonumber\\
&& \quad  \quad \, \quad\quad-\frac{1}{t^2}
 e^{i (\lambda+2 \theta+8 k \pi 
\beta-2\beta\pi)}    .\nonumber
\eea
Following Herman \cite{he}, we first get the positivity of the Lyapunov 
exponent.
\begin{prop}\label{aperiod} 
Let $T(k)^{\theta}$ be the
transfer matrices (\ref{tape}) at spectral parameter $\lambda\in \T$ 
corresponding to $U\equiv M(\{\theta_k\}, \{\alpha\},\{\gamma_k\})$, 
where the $\theta_k$'s are given by (\ref{alpe}). 
For $\beta$ irrational, the Lyapunov exponent 
$\gamma(\lambda)$ satisfies for almost all $\theta$:
$$\gamma(\lambda) \geq \ln \frac{1}{t^2} > 0 ,\,\,\mbox{  therefore  }
\,\,\sigma_{ac}(U) = \emptyset.$$
\end{prop}
{\bf Proof:} We first note that the sub-additive ergodic theorem 
applies to $F_N(\theta)=\ln \|\Pi_{k=1}^NT(k)^{\theta}\|$ and
since $\tau$ is ergodic, 
\be\label{set}
  \lim_{N\ra\infty}\frac{F_N(\theta)}{N}=\gamma(\lambda)
\ee
almost surely with respect to $\P_0$.
Setting $z = e^{-i \theta}$, we write our transfer matrices $T(k,z)$,
 expliciting the dependence in $z \in {\mathbb C}^*$, and we define 
three matrices $(R_j(k))$, $j=-2, 0, 2$, by
\be
T(k,z) \equiv z^{2} R_{2}(k) + R_{0}(k) + z^{-2} R_{-2}(k) \enspace ,
\ee
where 
\bea
&&R_{2}(k) = e^{-i (\lambda+8 k \pi \beta-6\beta\pi)}  
\left( \begin{array}{cc} -1 & \frac{ir}{t} \\
-\frac{ir}{t}  & -\frac{r^2}{t^2}
      \end{array} \right) \enspace ,\\
&& R_{0}(k)=  \left( \begin{array}{cc} 0 & -\frac{ir}{t}\\
\frac{ir}{t} e^{4i \pi \beta} &  \frac{r^2}{t^2} (e^{i4 \pi \beta}+1)
      \end{array} \right) \enspace,\enspace \\
&&R_{-2}(k) =  -\frac{1}{t^2}
 e^{i (\lambda+8 k \pi 
\beta-2\beta\pi)}  \left( \begin{array}{cc} 0 & 0\\
0 & 1
      \end{array} \right) \enspace .
\eea
Then we consider $S_k(z) = z^2 T(k,z)$ which is analytic in $\C$ and
such that if $z \in S^1$, $\|S_k(z)\|=
\|T_k(z)\|$,  $\forall k \in {\mathbb Z}$. Hence, the function
$\| \ln \prod_{k = 1}^{N}S_k (z) \|$ is sub-harmonic and as
$S_k(0)=R_{-2}$, we get the estimate
\begin{equation}\label{1}
\frac{1}{2 \pi} \int_0^{2 \pi} \ln \|\prod_{k = 1}^{N}
S_k (e^{i \theta}) \| d\theta \geq \ln \|\prod_{k = 1}^{N}
S_k (0) \| = N \ln \frac{1}{t^2} \enspace .
\end{equation}
We finally note that (\ref{set}) implies
\be \gamma(\lambda) = \lim_{N \rightarrow +\infty}
\frac{1}{N} \int_0^{2 \pi} \ln
\|\prod_{k=1}^{N} T_k(e^{i\theta})\| \frac{d \theta}{2 \pi}.
\ee
The second statement then follows from Theorem \ref{IP}.
\hfill \ep\\
Next, we adapt the argument of Gordon to our setting
in order to exclude eigenvalues in $\sigma(U)$, for 
$\beta$ a Liouville number. That is if for any $k\in\N$,
there exists $p_k,q_k\in\N$ such that
\be\label{Liouvil}
  |\beta-p_k/q_k|\leq k^{-q_k}.
\ee
\begin{prop}\label{Gordon}
Assume the phases $(\theta_k)$ are given by (\ref{alpe}) 
and $(\alpha_k)$ are zero.
Moreover, suppose $\beta$ is a Liouville number and 
$((\theta^m)_k)$ a family of
periodic sequence of period $q_m$. For each sequence, the
corresponding family of transfer matrices (\ref{tape}) is denoted by
$(T(k)^{\theta})_{k \in {\mathbb Z}}$ and $(T_{m}^{\theta}(k))_{k \in
  {\mathbb Z}}$ respectively. Assume the period of the sequence 
$(\theta^m)$
obeys $\lim_{m \rightarrow + \infty} q_m = +\infty$ and the
following estimates holds 
$$\sup_{k, m} \| T_{m}^{\theta}(k)\|<\infty,\,\,\,\,\,\,
\sup_{|k| \leq 2q_m} \|T^{\theta}(k) -T^{\theta}_{m}(k)\| \leq C m^{-q_m}.$$ 
Then, any non zero solution $\phi=\sum c_k\ffi_k$ of
$U\phi = e^{i\lambda}\phi$ satisfies
\be
\limsup_{|k|\rightarrow +\infty} \frac{c_{k+1}^2 +
  c_k^2}{c_1^2 + c_0^2} \geq \frac{1}{4} \enspace .
\ee 
\end{prop}
Its proof is identical to that given 
in \cite{cfks}, Theorem 10.3., noting that the norm of any transfer 
matrix (\ref{tape}) is bounded by a constant depending on $t,r$ only.
\begin{thm}
Let $U$ be as in Proposition \ref{aperiod}. If $\beta$ is a 
Liouville number, then for a.e. $\theta$, $U$ is purely singular 
continuous.
\end{thm}
\noindent {\bf Proof: } Let $\beta$ be a Liouville number. It can be
approximated by  a sequence of irreducible fractions
$(\frac{p_m}{q_m})$ obeying (\ref{Liouvil}). Define the sequence
$((\theta^m)_k)$ by: $\forall k \in {\mathbb Z}$, $(\theta^m)_k = 2\pi
\frac{p_m}{q_m} k + \theta$. A simple computation shows $\forall k 
\in {\mathbb Z}$
\bea\label{difference}
&&T^{\theta}(k)-T_{m}^{\theta}(k) = \\
&&2i \sin\left((4k-3)\pi(\beta-\frac{p_m}{q_m})\right)
e^{-i\left(\lambda+2\theta+(4k-3)\pi(\beta+\frac{p_m}{q_m})\right)}\left(\begin{array}{cc}1& -irt^{-1}\\irt^{-1}& r^2t^{-2}\end{array}\right)\nonumber\\
&& -2i \sin\left((4k-1)\pi(\beta-\frac{p_m}{q_m})\right)
e^{i\left(\lambda+2\theta+(4k-1)\pi(\beta+\frac{p_m}{q_m})\right)}\left(\begin{array}{cc}0&0\\ 0&t^{-2}\end{array}\right)\nonumber
\eea
Then, $\beta$ being a
Liouville number allows us to check the hypotheses of the Proposition
\ref{Gordon}. Therefore, the generalized eigenvalue equation cannot
have $l^2$ solution and the point spectrum of $U$ is
empty. Combining this result with Theorem \ref{aperiod} yields 
the conclusion.
\ep

\setcounter{equation}{0}
\section{Appendix}

{\bf Proof of Lemma \ref{trivtri}:}\\
Assume $U$ is unitary and , for all $k\in\Z$,
\be
U\ffi_k=\alpha_k\ffi_{k-1}+\beta_k\ffi_k+\gamma_k\ffi_{k+1},
\ee
so that
\be
U=\pmatrix{   \ddots &\alpha_{k-1} & & & \cr
                    &\beta_{k-1}&\alpha_k & &   \cr
                    & \gamma_{k-1}&\beta_k&\alpha_{k+1}&  \cr
                    & & \gamma_k&\beta_{k+1}& \cr    
                   & & &\gamma_{k+1} &  \ddots   }.
\ee
Then, , for all $k\in\Z$,
\bea\label{relation}
& &|\alpha_k|^2+|\beta_k|^2+|\gamma_k|^2=1 \nonumber\\
& &|\alpha_{k+1}|^2+|\beta_k|^2+|\gamma_{k-1}|^2=1 \nonumber\\
& &\alpha_k\beta_{k-1}^*+\beta_k\gamma_{k-1}^*=0 \nonumber\\
& &\gamma_{k-1}\beta_{k-1}^*+\beta_k\alpha_k^*=0\nonumber\\
& &\alpha_k\gamma_{k}^*=0\nonumber\\
& &\alpha_k\gamma_{k-2}^*=0
\eea

Let us start by noting that $|\beta_{k_0}|=1$ is equivalent 
to $\alpha_{k_0}=\gamma_{k_0}=\alpha_{k_0+1}=\gamma_{k_0-1}=0$,
which creates an isolated $1\times 1$ block in the matrix
structure of $U$.

We assume now that one off-diagonal element is non zero.
By considering the transpose of $U$ instead of $U$ if necessary,
we assume without loss $\alpha_{k_0}\neq 0$. The last two relations
impose $\gamma_{k_0}=\gamma_{k_0-2}=0$ and the two middle one
yield 
\be\label{above}
|\beta_{k_0}\alpha_{k_0+1}|=|\beta_{k_0-1}\alpha_{k_0-1}|=
|\beta_{k_0-1}\alpha_{k_0-1}|=|\beta_{k_0-2}\alpha_{k_0-1}|=0.
\ee
On the one hand, if $\beta_{k_0}\neq 0$, then $\beta_{k_0-1}\neq 0$. Otherwise
we would get from the first two relations in (\ref{relation}) 
$|\alpha_{k_0}|=1$ and $\beta_{k_0}= 0$. Hence, from
(\ref{above}),  $\alpha_{k_0-1}=\alpha_{k_0+1}=0$ showing 
that an isolated bloc of the form
\be
\pmatrix{\beta_{k_0-1}& \alpha_{k_0}\cr\gamma_{k_0-1}& \beta_{k_0} }
\ee
exists in the matrix $U$. 

If, on the other hand, $\beta_{k_0}=0$, together with $\alpha_{k_0}\neq 0$
this implies $|\alpha_{k_0}|=1$ and, in turn
$\beta_{k_0-1}=0$. 

We first assume $\gamma_{k_0-1}\neq 0$. Hence the last two 
equations in (\ref{relation}) yield $\alpha_{k_0-1}=\alpha_{k_0+1}=0$,
which again yields an isolated block of the form 
\be
\pmatrix{0& \alpha_{k_0}\cr\gamma_{k_0-1}&0 }
\ee
in $U$ with $|\gamma_{k_0-1}|=|\alpha_{k_0}|=1$. 

If $\gamma_{k_0-1}=0$ and $|\alpha_{k_0}|=1$, we get
$|\alpha_{k_0-1}|=|\alpha_{k_0+1}|=1$. In turn, this 
imposes $\beta_{k_0-1}=\gamma_{k_0-1}=0$ and 
$\beta_{k_0+1}=\gamma_{k_0+1}=0$. Thus,
$U$ is of the form
\be
  U=\pmatrix{   \ddots &\alpha_{k_0-1} & & & \cr
                   &0 &\alpha_{k_0} & &   \cr
                   & &0 &\alpha_{k_0+1} &  \cr
                   && &0 & \cr    
                  & & & & \ddots     }
\ee
and is therefore unitarily equivalent to the shift operator, using 
a unitary defined similarly to (\ref{unit}). 

Hence, except in the last case, iteration of the above arguments, 
shows that $U$ has the block structure announced. \hfill\ep\\

\vspace{.2cm}
\noindent{\bf Proof of Lemma \ref{indep}:} We can set the value $\Lambda$ at
zero without loss. Let 
\be\label{charf}
\Phi_{u}(n)={\mathbb E}(e^{-in\theta})=\delta_{n,0}
\ee
be the characteristic function of the common uniform distribution
of the phases $\theta_k$ and $\alpha_k$. 
Consider the characteristic function of the set 
of random vectors
$\{\delta_{k_1}, \delta _{k_2}, \cdots \delta _{k_j}\}$ given by
\bea
&&  \Phi_{ \delta _{k_1}, \delta _{k_2}, \cdots \delta _{k_j}}
(n_1, n_2, \cdots, n_j)= {\mathbb E}(\exp(-i(n_1\cdot \delta_{k_1}+
n_2\cdot \delta_{k_2}+\cdots + n_j\cdot \delta_{k_j} )))\nonumber\\
&& = {\mathbb E}(\exp(-i( n_1^1 \theta_{2k_1}+(n_1^1+n_1^2)
\theta_{2k_1-1}+n_1^2\theta_{2k_1-2}+\cdots +
n_j^1 \theta_{2k_j}\nonumber\\
&&\,\,\,\,\,\,\,\,\,\,\,\,\,\,\,\,\,\,\,\,\,\,\,\,\,\,\,\,\,\,
+(n_j^1+n_j^2)\theta_{2k_j-1}+n_j^2\theta_{2k_j-2})))\times
\nonumber\\
&&{\mathbb E}(\exp(-i( n_1^1 \alpha_{2k_1}+(n_1^2-n_1^1)
\alpha_{2k_1-1}-n_1^2\alpha_{2k_1-2}+\cdots+
n_j^1 \alpha_{2k_j}\nonumber\\
&& \,\,\,\,\,\,\,\,\,\,\,\,\,\,\,\,\,\,\,\,\,\,\,\,\,\,\,\,\,\,
+(n_j^2-n_j^1)\alpha_{2k_j-1}-n_j^2\alpha_{2k_j-2}))).
\eea
where $n_k=(n_k^1, n_k^2)\in\Z^2$.  We used independence of the 
$\theta$'s and $\alpha$'s to factorize the expectations over these 
random variables. We can assume the $k_l$'s are ordered and we deal 
with the $\theta$'s only. The argument is similar for the $\alpha$'s. 
From the expression above, one sees that one can factorize the 
expectations over $\theta_{l}$ with $l\leq 2k_r$ from those with
$l\geq 2k_{r+1}-2$ as soon as $k_r<k_{r+1}+1$. Therefore, it is
enough to consider consecutive indices $k_1=m, k_2=m+1, \cdots, k_j=m+j$.
As (\ref{charf}) shows, in such  a case, the expectation over the 
$\theta$'s equals zero unless
\be
  n_1^1=0, \,\,\,\, n_1^1+n_1^2=0, \,\,\,\, \cdots \,\,\,\, , n_j^1+n_j^2=0, \,\,\,\, n_j^2=0,
\ee
when it equals one. But this is equivalent to $n_k^{l}=0$ for all
$k=1, \cdots, j$, $l=1,2$. Hence, we have proven that
\be
  \Phi_{\delta}(n)=\Phi_{u}(n^1)\Phi_{u}(n^2)
\ee
for $j=1$ and
\be
\Phi_{ \delta _{k_1}, \delta _{k_2}, \cdots \delta _{k_j}}
(n_1, n_2, \cdots, n_j)=\Phi_{\delta _{k_1}}(n_1)
\Phi_{\delta _{k_2}}(n_2)\cdots \Phi_{\delta _{k_j}}(n_j),
\ee
which is equivalent to independence of the random vectors
$\delta _{k_1}, \delta _{k_2}, \cdots \delta _{k_j}$. \ep

\vspace{.2cm}
\noindent{\bf Proof of Lemma \ref{shortproof}:} Let us consider 
$Y_k=Y_k^+=X_k+X_{k-1}$ only, the other case being similar. Let
the measure $\mu_X$ denote the distribution of the $X_k$'s. Then
the $Y_k$'s are identically distributed according to the measure 
$\mu_{Y}=\mu_X*\mu_X$. Let $\Phi_X$ be the characteristic function
of the random variable $X$.
Then $\Phi_Y(n)= \Phi_X^2(n)$. Given Lemma \ref{indep}, we need only prove that independence of
the $Y_k$'s imposes $\mu_X$ is uniform on the torus. 
Then, the characteristic function of the variables 
$\{Y_{k}, Y_{k+1}\}$ must satisfy for all $(n_1, n_2)\in \Z^2$
\bea
  &&\Phi_{ Y_{k}, Y_{k+1}}(n_1, n_2)={\mathbb E} (\exp(-in_1(X_{k}+
X_{k -1})-in_2(X_{k+1}+X_{k})))\nonumber\\
&&= \Phi_X(n_1)\Phi_X(n_1+n_2)\Phi_X(n_2)\equiv \Phi_X^2(n_1)\Phi_X^2(n_2).
\eea
In case $\Phi_X(n_1) \Phi_X(n_2)=0$, this relation is fulfilled. Otherwise,
we have for all other cases 
\be\label{rule}
\Phi_X(n_1) \Phi_X(n_2)=\Phi_X(n_1+n_2).
\ee
If $N$ is the smallest positive integer such that $\Phi_X (N)\neq 0$, we get that
\be
1=\Phi_X (0)=\Phi_X (N)\Phi_X (-N)=|\Phi_X (N)|^2\,\, \Longleftrightarrow
\Phi_X (N)=\E^{-i\nu},
\ee
for some $\nu\in\T$. Iteration of (\ref{rule}) implies that for any $m\in\Z$,
\be
 \Phi_X (mN)=\E^{-im\nu}.
\ee
That implies that $\mu_x=\delta(x-\nu/N)$, which is a contradiction to our
hypothesis. Hence we must have $\Phi_x(n)=0$ for all $n\neq 0$, which 
corresponds to a uniform distribution. \hfill\ep\\

\vspace{.2cm}
\noindent{\bf Proof of Theorem \ref{solpoly}:}
We develop here the arguments yielding polynomially bounded
generalized eigenfunctions associated with spectral parameters 
in the spectrum of $U$. We state the starting point result, Theorem 
C.5.1 in \cite{s}, specialized to our setting.
\begin{thm}\label{radonnyk}
Let ${\cal H}$ be a separable Hilbert space. Assume that to any Borel set 
$\Delta \subset [0,2\pi[$ we have a positive trace class operator 
  $A(\Delta)$ on ${\cal H}$ satisfying: the condition
 if $\Delta = \cup_{n=1}^{+\infty} \Delta_n$ with  $\Delta_i \cap
  \Delta_j =\emptyset$ for  $i\neq j$, then $A(\Delta) = s-\lim \sum
  A(\Delta_n)$. 

Then there exists a Borel measure $d\rho$ and a positive, trace class, 
operator valued measurable function $a(\lambda)$ such that:
\begin{itemize}
\item $\forall \phi \in {\cal H}$, $\bra \phi | A(\Delta) \phi
  \ket = \int_{\Delta} \bra \phi | a(\lambda) \phi \ket
  d\rho(\lambda)$
\item $Tr(a(\lambda)) = 1$, $d\rho$-ae.
\end{itemize}
These two conditions characterize the operator valued function $a$.
\end{thm}
Let us introduce weighted $l^2(\Z)$ spaces
\be
l^2_{\delta}({\mathbb Z}) = \{ \phi = (\phi_n)_n \in
  l^2({\mathbb Z}^*); \sum_{n \in {\mathbb Z}} <n>^{\delta} 
|\phi_n|^2 < +\infty  \} ,
\ee
where $<n>=<1+n^2>^{1/2}$.
We prove the equivalent of Theorem C.5.2. in \cite{s}.
\begin{prop} \label{intermed}
Let $U$ be a unitary operator defined on $l^2({\mathbb Z})$ and $\delta > 1$. 
Then there exists a spectral measure $d\rho$ and, 
for $d\rho$ almost all $\lambda$, there exists a
  function $F_{.,.}(\lambda)$ defined on ${\mathbb Z} \times {\mathbb
  Z}$ such that:
\begin{itemize}
\item $F_{n,m}$ is measurable in $\lambda$,
\item $\sum_{n,m} <n>^{-\delta} |F_{n,m}(\lambda)|^2 <m>^{-\delta} 
\leq 1$, $d\rho$-ae
\item $|F_{n,m}(\lambda)| \leq C <n>^{-\frac{\delta}{2}}
  <m>^{-\frac{\delta}{2}}$
\item For any bounded Borel function $g$ on $S^1$, and for any
  vectors $\phi$, $\psi$  in $l^2_{\delta}({\mathbb Z})$,
\be
\bra \phi | g(U) \psi \ket = \int g(\lambda) \left(\sum_{n,m} F_{n,m}(\lambda)
{\phi_n}^* \psi_m \right) d\rho(\lambda).
\ee
\item For any fixed $m$, $(U- e^{i\lambda}) F_{.,m}(\lambda) = 0$,
where  $F_{.,m}(\lambda)=\sum_{n\in\Z}F_{n,m}(\lambda)\ffi_n$.
\end{itemize}
\end{prop}
{\bf Proof: } We denote the spectral projectors of $U$
by $(E(\Delta))_{\Delta \in {\cal B}([0,2\pi[)}$ where ${\cal
B}([0,2\pi[)$ denotes the Borel sets on the interval $[0,2\pi[$. 
Let $x$ be the self adjoint operator, diagonal on the orthonormal basis 
$(\ffi_n)_{n \in {\mathbb Z}}$, defined  $\forall n \in {\mathbb Z}$, 
by $x\ffi_n = <n>\ffi_n$. 
The operators  $(A(\Delta))_{\Delta \in {\cal B}([0,2\pi[)}$ defined 
$\forall \Delta \in {\cal B}([0,2\pi[)$ by
\be
 A(\Delta)=
x^{\frac{-\delta}{2}}E(\Delta)x^{\frac{-\delta}{2}} \enspace,
\ee
are positive and trace class:
\be
\sum_{n \in {\mathbb Z}} \bra \ffi_n |
x^{\frac{-\delta}{2}}E(\Delta)x^{\frac{-\delta}{2}} \ffi_n \ket \leq
\sum_{n \in {\mathbb Z}} <n>^{-\delta} \bra \ffi_n | E(\Delta) \ffi_n
  \ket < +\infty  .
\ee
By definition, the spectral family $E(.)$ satisfies for any countable 
disjoint family $(\Delta_i)_{i \in I} \subset {\cal B}([0,2\pi[)$: $E(\cup_{i
  \in I} \Delta_i) = s-\lim \sum_{i \in I} E(\Delta_i)$. The
  operators $x^{-\delta}$ being bounded on $l^2({\mathbb
  Z})$, we get $A(\cup_{i \in I} \Delta_i) = s-\lim \sum_{i \in I} 
A(\Delta_i)$.
Hence $A(.)$ is a Borel measure with values in positive, trace class 
operators. 
and Theorem \ref{radonnyk} applies. Thus, $\forall (n,m) \in {\mathbb
  Z} \times {\mathbb Z}$, we get a function defined $d\rho$-ae 
\bea
 F_{n,m}(\lambda) &=& \bra \ffi_n | x^{\frac{\delta}{2}} a(\lambda)
 x^{\frac{\delta}{2}} \ffi_m \ket = (<n><m>)^{\frac{\delta}{2}} \bra \ffi_n
 |a(\lambda) \ffi_m \ket \nonumber\\
&=& (<n><m>)^{\frac{\delta}{2}} a_{n,m}(\lambda).
\eea
By construction, the functions $a_{n,m}$ (hence $F_{n,m}$) are 
measurable. Moreover,
\bea
&&\sum_{n,m} |F_{n,m}(\lambda)|^2 (<n><m>)^{-\delta} = \sum_{n,m}
|a_{n,m}(\lambda)|^2\\
\quad\quad&&= \sum_n \|a(\lambda)\ffi_n\|^2 
= \|a(\lambda)\|_2^2
 \leq \|a(\lambda)\|_1^2 = Tr(a(\lambda))^2 =1 \enspace d\rho-ae
\enspace .\nonumber
\eea
This implies the third statement. 
Let $A_i \subset [0,2\pi[$ be Borel set,
$\chi_i: S^1\ra \R$ be its characteristic function and 
$\phi, \psi$ be two vectors
of $l^2_{\delta}({\mathbb Z})$. Then
\bea
&&\int_{[0,2\pi[} \chi_i(e^{i\lambda}) \sum_{n,m} \phi_n^* \psi_m
F_{n,m}(\lambda) d\rho(\lambda) = \sum_{n,m} \phi_n^* \psi_m
\int_{[0,2\pi[} \chi_i(e^{i\lambda})F_{n,m}(\lambda) d\rho(\lambda)\nonumber\\
&& = \sum_{n,m} \phi_n^* \psi_m
\int_{A_i} F_{n,m}(\lambda) d\rho(\lambda)
= \sum_{n,m} \phi_n^* \psi_m \bra \ffi_n | E(A_i) \ffi_m \ket \nonumber\\
&&= \sum_{n,m} \phi_n^* \psi_m \bra \ffi_n | (\int_{[0,2\pi[}
\chi_i(e^{i\lambda}) d\lambda)  \ffi_m \ket
= \bra \phi | \chi_i(U) \psi \ket \enspace .
\eea
This results holds for step functions by linearity, and for 
bounded measurable functions on $[0,2\pi[$. 
In particular, taking $g=id$ and $\psi =\ffi_m$,
\bea
\bra \phi| U \ffi_m \ket &=& \int_{[0,2\pi[} e^{i\lambda} \left( \sum_{n}
  F_{n,m} (\lambda) \phi_n^* \right) d\rho(\lambda) \enspace,\nonumber\\
&=& \int_{[0,2\pi[} \bra \phi | e^{i\lambda} F_{.,m}(\lambda) \ket 
d\rho(\lambda) .
\eea
But,
\bea
&& \int_{[0,2\pi[} \bra \phi | UF_{.,m}(\lambda) \ket d\rho(\lambda) =
\sum_{k} \int_{[0,2\pi[} \phi_k^* (UF_{.,m}(\lambda))_k d\rho(\lambda)\nonumber\\
&&= \sum_{k} \int_{[0,2\pi[} \phi_k^* \sum_{j} U_{kj} F_{j,m}(\lambda)
d\rho(\lambda)
= \sum_{k,j} U_{kj} \phi_k^* \int_{[0,2\pi[} F_{j,m}(\lambda)
d\rho(\lambda)\nonumber\\
&&= \sum_{k,j} U_{kj} \phi_k^* (jm)^{\delta} \int_{[0,2\pi[} a_{j,m}(\lambda)
d\rho(\lambda)
= \sum_{k,j} U_{kj} \phi_k^* (jm)^{\delta} A([0,2\pi[)_{j,m}\nonumber\\
&&= \sum_{k,j} U_{kj} \phi_k^* E([0,2\pi[)_{j,m} = \sum_{k,j} U_{kj} 
\phi_k^* {\delta}_{j,m}= \bra \phi | U\ffi_m \ket .
\eea It follows that $\forall m \in {\mathbb Z}, \forall \phi \in
l^2_{\delta}({\mathbb Z})$, 
\be
\int_{[0,2\pi[} \bra \phi | UF_{.,m}(\lambda) \ket d\rho(\lambda) =
\int_{[0,2\pi[} \bra \phi | e^{i\lambda} F_{.,m}(\lambda) \ket d \rho,
\ee and thus
\be
\bra \phi | UF_{.,m}(\lambda) \ket = \bra \phi | e^{i\lambda} F_{.,m}(\lambda) \ket
\enspace d\rho-ae \enspace .
\ee
\ep 

At this point we can prove Theorem \ref{solpoly}, following closely 
the arguments of \cite{s}:
Let $N(\lambda)$ be the rank of the  Hilbert-Schmidt operator $a(\lambda)$,
which is a measurable function of $\lambda$. For all 
$\lambda$, there exists a set of orthogonal 
vectors \cite{k},
$(f_j(\lambda))_{j \in \{1, \ldots ,N(\lambda)\}}$, such that 
d$\rho$-ae:
\bea
&&a(\lambda) = \sum_{j=1}^{N(\lambda)} |f_j(\lambda) \ket \bra
f_j(\lambda)| \,\,\,\,\mbox{ and }\nonumber\\
&&\sum_{j=1}^{N(\lambda)} \|f_j(\lambda)\|^2 = \sum_{m,j}
\frac{1}{\|f_m(\lambda)\|^2} \bra f_m(\lambda)| f_j(\lambda) \ket \bra
f_j(\lambda) | f_m(\lambda) \ket \nonumber\\&&=\sum_{m=1}^{N(\lambda)} \bra
\frac{f_m(\lambda)}{\|f_m(\lambda)\|}| a(\lambda)
\frac{f_m(\lambda)}{\|f_m(\lambda)\|}\ket = Tr(a(\lambda)) = 1\enspace .\nonumber
\eea
In case of degeneracy of the spectrum, it is always possible \cite{s}
to chose the  $f$'s so that they are measurable. It is enough to set now
\be
\phi_n(\lambda) = x^{\delta/2} f_n(\lambda) \enspace ,
\forall n \in {\mathbb Z}, \forall \lambda \in [0,2\pi[, \enspace \Delta_n
= \{\lambda; N(\lambda) = n\} \enspace .
\ee
The sets $\Delta_n$ are disjoint by construction. 
For any fixed  $\lambda$, the vectors $\phi_j(\lambda)$ are linearly 
independent, as easily checked.
The conditions on the growth of the components of the vectors 
$\phi_j(\lambda)$ are consequences of their definitions and Proposition
\ref{intermed}.
By construction, $\forall k \in {\mathbb Z}$,
\be
\|f_j(\lambda)\|^2 (\phi_j(\lambda))_k = \sum_{m} <m>^{-\delta}
F_{k,m}(\lambda) (\phi_j(\lambda))_m
\ee
Therefore, $\forall n \in {\mathbb Z}, \forall j \in \{1, \ldots , N(\lambda)\}$,
\bea
\bra \ffi_n| U \phi_j(\lambda) \ket &=&
\sum_{k} U_{nk} (\phi_j(\lambda))_k \nonumber\\
&=& \frac{1}{\|f_j(\lambda)\|^2} \sum_{k,m} U_{nk} <m>^{-\delta}
F_{k,m}(\lambda) (\phi_j(\lambda))_m \nonumber\\
&=& \frac{1}{\|f_j(\lambda)\|^2} \sum_{m} <m>^{-\delta}
(\phi_j(\lambda))_m \bra \ffi_n | UF_{.,m}(\lambda)\ket \nonumber
\eea
Using Proposition \ref{intermed}, it follows that the previous line equals
\be
= \frac{1}{\|f_j(\lambda)\|^2} \sum_{m} <m>^{-\delta}
(\phi_j(\lambda))_m e^{i\lambda} \bra \ffi_n | F_{.,m}(\lambda)\ket
= \bra \ffi_n| e^{i\lambda} \phi_j(\lambda) \ket  .
\ee
Thus, $\forall \phi \in l^2_{\delta}({\mathbb Z})$, $\bra \phi| U
\phi_j(\lambda) \ket = \bra \phi | e^{i\lambda} \phi_j(\lambda) \ket$, 
d$\rho$-ae.
\ep

\vspace{.2cm}
\noindent{\bf Proof of Lemma \ref{sqrt}:} Write 
\be
 R(\lambda)=\pmatrix{a(\lambda) & b(\lambda)\cr c(\lambda)  & d(\lambda)  }
\ee
where $a, b, c, d$ are analytic on $\T$ and $\det R(\lambda)=e^{i\kappa} $. 
The eigenvalues of $R(\lambda)$ are
\be
  E_{j}(\lambda)=\frac{\mbox{Tr}R(\lambda)}{2}+
(-1)^{j}\sqrt{\frac{(\mbox{Tr}R(\lambda))^2}{4}-e^{i\kappa} }\,\,\, j=1,2
\ee
and the set $X$ consists of the zeros
of $(\mbox{Tr}R)^2-4e^{i\kappa}$. 
Let $\lambda=0$ belongs to $X$. We can assume that in a punctured 
neighborhood of $0$,
$b(\lambda)\neq 0$. Therefore, the eigenvectors can be chosen as
\be
v_j(\lambda)=\pmatrix{b(\lambda)\cr E_j(\lambda)-a(\lambda)}.
\ee
Since 
\be
(\mbox{Tr}R(\lambda))^2/4-e^{i\kappa}=\sum_{n\in\N}t_n\lambda^{n}
\ee
with $t_0=0$, $E_j$ and, in turn, $v_j$ admit convergent series 
expansions in non-negative powers of $\lambda^{1/2}$ in a neighborhood of
$0$. \ep

\vspace{.5cm}

{\bf Acknowledgements:}\\
We wish to thank Joachim Asch and Jean Brossard for helpful discussions.
JH wishes to thank the Institut Fourier and AJ wishes to thank the University of Virginia for hospitality and support.

\end{document}